\documentclass[aps,pra,epsf,superscriptaddress,amsmath,amssymb,amsfonts,twocolumn,showpacs,nofootinbib]{revtex4-1}

\usepackage{graphicx}
\usepackage{dcolumn}
\usepackage{bm}
\usepackage{braket}
\usepackage{amsmath}
\usepackage{mathtools}
\usepackage{graphicx,color,xcolor}
\usepackage{hyperref}
\hypersetup{colorlinks=true}
\usepackage[capitalize]{cleveref}
\newcommand{\abs}[1]{\left| #1 \right|} 
\usepackage{multirow}

\usepackage[normalem]{ulem} 

\usepackage[sectionbib]{bibunits}
\usepackage{bibunits}

\usepackage{array}
\usepackage{pict2e}
\usepackage{pifont}

\begin{document}

\begin{bibunit}[apsrev4-1]

\title{Observation of vector rogue waves in repulsive three-component atomic mixtures}

\author{G. A. Bougas}
\email{gbougas@mst.edu}
\affiliation{Department of Physics and LAMOR, Missouri University of Science and Technology, Rolla, MO 65409, USA}

\author{G. C. Katsimiga}
\affiliation{Department of Physics and LAMOR, Missouri University of Science and Technology, Rolla, MO 65409, USA}%

\author{S. Mossman}
\affiliation{Department of Physics and Biophysics, University of San Diego, San Diego, California 92110, USA}
\affiliation{Department of Physics and Astronomy, Washington State University, Pullman, Washington 99164-2814, USA}

\author{P. Engels}
\affiliation{Department of Physics and Astronomy, Washington State University, Pullman, Washington 99164-2814, USA}

\author{P. G. Kevrekidis}
\affiliation{Department of Mathematics and Statistics, University of Massachusetts Amherst, Amherst, MA 01003-4515, USA}%

\affiliation{Department of Physics, University of Massachusetts Amherst, Amherst, MA 01003-4515, USA}
\affiliation{Theoretical Sciences Visiting Program, Okinawa Institute of Science and Technology Graduate University, Onna, 904-0495, Japan}

\author{S. I. Mistakidis}
\affiliation{Department of Physics and LAMOR, Missouri University of Science and Technology, Rolla, MO 65409, USA}

\date{\today}

\begin{abstract} 
Rogue waves are extreme evanescent nonlinear structures that are challenging to observe in atomic gases, as their emergence requires a dynamically unstable attractive environment.
Here, we report the experimental observation of vector 
extensions of Peregrine solitons in highly particle-imbalanced, pairwise immiscible three-component repulsive Bose-Einstein condensates. 
The possibility of an effectively attractive character of the minority components is established by constructing a generalized reduction scheme for an imbalanced $\mathcal{N}$-component
setup with arbitrary interaction signs. 
These components 
are subject to
intra- and inter-component modulation instability, which along with the presence of an attractive potential well induces the dynamical formation of highly reproducible vector rogue waves. 
Exploiting different Rb hyperfine states, it is possible to flexibly 
tune the effective interactions stimulating the realization of 
a plethora of vector rogue waves, including single and double Peregrine-like wave peaks. 
The experimental findings are in quantitative agreement with suitable three-dimensional mean-field  simulations, 
while quasi-one-dimensional analysis 
of the non-polynomial Schr{\"o}dinger model provides additional insights into the rogue wave characteristics. 
\end{abstract}

\maketitle

{\vspace{0.25cm}\noindent\large\textbf{Introduction}}
\label{Sec:Intro}

Rogue waves (RWs), originally measured through the Draupner platform in North sea~\cite{kharif2008rogue}, are extreme nonlinear wave events of great steepness arising suddenly and dissolving without a trace~\cite{hopkin2004sea,kharif2008rogue}. 
Based on their oceanographic definition~\cite{Fu_review_2026}, RWs are highly localized wave structures, whose 
amplitude
exceeds at least twice the average background height.
These  waves are ubiquitous in nature and are observed in fields ranging from water tanks~\cite{Chabchoub_water,Chabchoub_water1}, plasmas~\cite{Bailung_plasmas}, nonlinear optics~\cite{kibler2010peregrine,solli2007optical}, to oceanography~\cite{kharif2008rogue,perkins2006dashing}, atmosphere~\cite{stenflo2010rogue}, capillaries~\cite{Shats_RWs} and even argued to arise in financial markets~\cite{yan2010financial,yan2011vector}. 
Prototypes of RWs, within the generic
nonlinear Schr{\"o}dinger model, include the Peregrine soliton (PS)~\cite{peregrine_water_1983} being localized in both space and time, as well as the Akhmediev~\cite{akhmediev2009waves} and Kuznetzov-Ma~\cite{kuznetsov1977solitons,ma1979perturbed} solitons that are periodic in space and time respectively.  
Note that the PS corresponds to the limit of infinite temporal and spatial periodicity of the Kuznetzov-Ma and Akhmediev breathers respectively.
A fundamental condition~\cite{ruderman2010freak,baronio_vector_2014}
accompanying RW formation is the exponential growth of periodic perturbations associated with modulational instability (MI) on an attractively interacting unstable uniform background~\cite{zakharov_modulation_2009,Everitt_observation_2017}.

Bose-Einstein condensates (BECs) offer highly tunable and accessible platforms 
to realize 
RWs and MI owing to their phenomenal controllability and handling of system parameters~\cite{bloch2008many,bloch2012quantum,mistakidis2023few}.   
For instance, MI has been exploited for generating bright solitary waves~\cite{strecker2002formation,strecker2003bright,nguyen2017formation}, the renowned Townes solitons~\cite{Chen_observation_2020} and necklaces thereof~\cite{banerjee2024collapse} as well as dispersive shock-waves~\cite{tamura2025observation} in genuinely attractive, lower dimensional, single-component BECs.  
Recently, immiscible binary BECs featuring repulsive interactions have been utilized not only to demonstrate the nonlinear stage of MI~\cite{mantzavinos} in particle balanced mixtures~\cite{mossman2024nonlinear} but also to  controllably nucleate the Townes soliton~\cite{Bakkali_realization_2021,bakkali_townes_2022} and PS~\cite{Romero_experimental_2024} in particle-imbalanced ones.  
The latter remarkable and unexpected features for fully repulsive media, are rooted in the reduction of a two-component immiscible system to an effectively attractive single-component one~\cite{Dutton_eff}. This avoids the complications stemming from wave collapse of an actual attractive condensate~\cite{Timmermans_phase_1998,Bakkali_realization_2021}.  

The versatility of BECs increases further when considering higher-component mixtures, such as three-component systems,~\cite{stamper2013spinor,Wu_threecomp,Taglieber_threecomp,Bersano,Lannig} which can support a wider range of phases and nonlinear excitations. 
Following up on corresponding observations of
one-component dark~\cite{Burger_darks,Weller_darks} 
and two-component dark-bright~\cite{becker2008oscillations} states,
spinor condensates have enabled the realization of
vector dark-bright-bright~\cite{Bersano,Lannig} solitons, magnetic solitons~\cite{Farolfi_observation_2020,Chai_magnetic_2020,Chai_magnetic_2021}, as well
as more complex patterns including ferrodark
(and antidark) solitary waves~\cite{ferrodark}.
However, an observation of vector RWs and their potential interactions remains 
elusive
in the cold atom realm, in contrast to realizations of dark 
multi-component RWs in
optics~\cite{Frisquet2016_SciRep_OpticalDarkRogueWave,kibler1}.
Here, vector RWs refer to the presence of a RW structure per component of the mixture 
~\cite{baronio_2012,Bludov_vector_2010,Guo_rogue_2011}.
Generation of such vector RWs demands exquisite controllability of the 
cold atom setting.
Additionally, the generalization of the above reduction scheme to $\mathcal{N}$-component immiscible repulsive  highly-imbalanced BECs to effective lower component attractive ones remains elusive. 
This paves the way for creating on demand effectively attractive multi-component BECs with tunable interactions.

In this work, we experimentally and theoretically showcase a prototype, 
in the context of BECs,
of the dynamical generation of vector RWs in the form of PS structures.
The latter can be reasonably fitted with the exact (in the integrable limit) PS waveform~\cite{Peregrine1983}. The PS is doubly localized in space and time, characterized by side density nodes, having a $\pi$ phase jump.
It is also important to recall such waveforms
are fundamentally distinct, e.g., from two-soliton 
waveforms whose time-periodic character has been
recently explored experimentally in~\cite{brochier2026breathingfissionmagneticmultisolitons}.

Here, the nucleation of vector RWs is achieved
in a highly controllable and  reproducible fashion, by exploiting a weak attractive potential well atop three-component, highly 
particle-imbalanced, repulsive BECs. 
The majority component is pairwise immiscible with at least one minority. Hence, the triple mixture can be reduced to a two-component BEC featuring a form
of  effectively attractive interactions.
Accordingly, it is able to host 
either  
i) single PSs in one minority component, or ii) vector PSs, i.e., one per minority component, as well
as iii) [vector] twin Peregrine structures, among others.
Here, the term twin PS refers to two doubly localized PS structures occurring in the same component, located at the two opposite sides of the trap center.
Additionally, vector RWs entail the formation of a RW per 
component of the effective two-component mixture, i.e. per minority component of the original setup.

To explain the presence of effective attractive interactions we {\it systematically} extract an analytical reduction scheme for a repulsive highly-imbalanced $\mathcal{N}$-component setting. 
This can lead to repulsive or attractive effective interactions, 
and provides a detailed understanding 
of our model while representing a generalization of the simpler earlier reduction (from two- to one-component) in Refs.~\cite{Dutton_eff,Bakkali_realization_2021,Romero_experimental_2024}. 
Equipped with the pairwise miscibility or immiscibility of the triple mixture components, along with the systematic analysis of the 
intra- and inter-species MI~\cite{kasamatsu_multiple_2004,kasamatsu_modulation_2006} of the reduced setting, we can a priori predict the vector RW nucleation. 
Our experimental measurements are in quantitative agreement with three-dimensional (3D) mean-field simulations in the presence of three-body losses. 
Extending the approach of Ref.~\cite{salasnich2009generalized}, we additionally construct a corresponding quasi-1D non-polynomial Gross-Pitaevskii equation (NPGPE), whose predictions are also found to be in agreement with the experimental findings. 
Within this framework we demonstrate a plethora of different vector PSs in various three-component settings.

Earlier two-component focusing~\cite{baronio_2012}
and defocusing~\cite{baronio_vector_2014}
nonlinear Schr{\"o}dinger models have been considered in the
famous Manakov limit in the context of RWs. On the other
hand, our study
provides a 
suitable
experimental platform and 
further motivation for theoretically and computationally 
studying the potential formation of
such RWs in a far broader {\it non-integrable}
ultracold atomic multi-component realm.
This, in turn, showcases the spinor BEC system as an ideal playground
for the exploration of multi-component RW patterns.

{\vspace{0.25cm}\noindent\large\textbf{Results}} 
\label{Sec:Rogues}

\begin{figure*}[t!]
\centering
\includegraphics[width=1\textwidth]{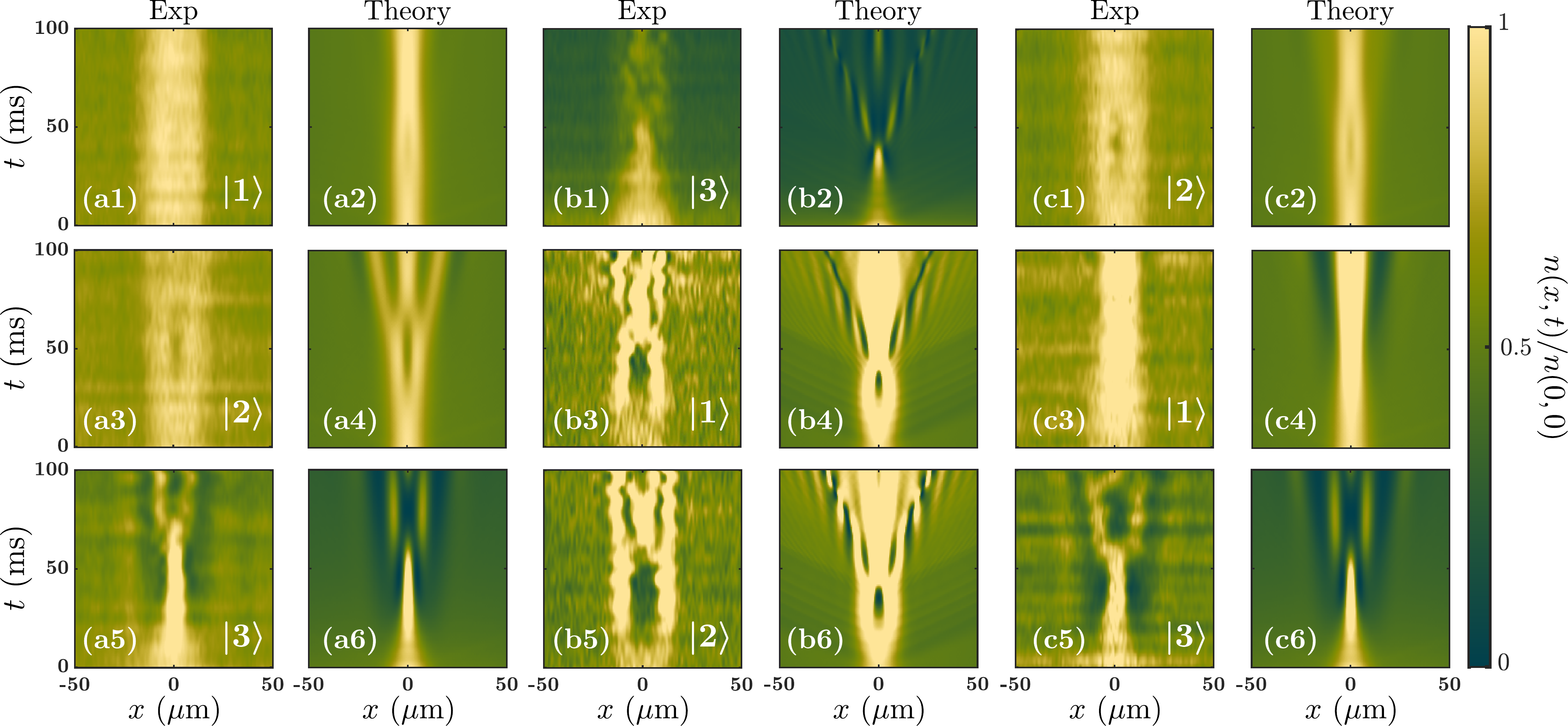}
\caption{\textit{{\bf Observation of vector PS configurations.}}  
Spatiotemporal evolution of the integrated absorption images [(a1), (a3), (a5); (b1), (b3), (b5); (c1), (c3), (c5)] is presented, averaged over 15 independent experimental realizations. Corresponding density distributions [(a2), (a4), (a6); (b2), (b4), (b6); (c2), (c4), (c6)] are obtained from 3D mean-field simulations. 
The setups pertain to (a1)-(a6) S1 , (b1)-(b6) S2 and (c1)-(c6) S3 consisting of $\ket{1} \equiv \ket{1,-1}$, $\ket{2} \equiv \ket{1,0}$ and $\ket{3} \equiv \ket{2,0}$ hyperfine states of $^{87}$Rb (see experimental panels). 
In S1 and S3 (S2) the majority-minority population imbalance reads $f_{\rm{m}}=10\% ~ (15\%)$. A PS forms in $\ket{3}$ followed by a density dip in $\ket{2}$ state for S1 and S3 respectively, while twin PS structures build upon both $\ket{1}$ and $\ket{2}$ accompanied by a PS in $\ket{3}$ in S2. Excellent agreement between the experiment and the 3D computations can be readily seen.}
\label{Fig:Spacetimes}
\end{figure*}

\textbf{Experimental setup.} 
We begin by preparing a single-component $^{87}$Rb BEC with approximately $N= 9\times 10^5$~atoms in an elongated optical trap with confinement frequencies $\omega = 2\pi \times (2.5, 246, 261)$~Hz.
The condensate is prepared in the $\lvert F, m_F\rangle = \lvert 1, -1\rangle \equiv \ket{1}$ state in the presence of a 10 G magnetic bias field which produces a sufficient quadratic Zeeman shift to make the various hyperfine states within the $F=1$ and $F=2$ manifolds individually addressable. Additionally, a 850 nm optical beam crosses the optical trap at a perpendicular angle along the vertical direction -- creating a weakly attractive well at the center of the condensate. 
The attractive well has a Gaussian shape of $w_x=12.6$ $\mu$m along the condensate's long axis and $w_y = 23.8$ $\mu$m perpendicular to the condensate with a potential depth of $V_0=39$ nK.  
Evaporating directly into this trap configuration suppresses bulk excitations and enhances reproducibility of the experimental procedure, see also the Methods
for further details.

Fast radio frequency and microwave pulses are subsequently employed to transfer population to the $\ket{1,0} \equiv \ket{2}$ and $\ket{2,0}\equiv \ket{3}$ states. Three distinct mixtures are created considering one of the $\ket{1}$, $\ket{2}$, $\ket{3}$ hyperfine levels as the majority component [M], while the remaining states serve as equally populated minority [m] components.
Explicitly, the three different setups read: 
\begin{subequations}
\begin{gather}
    \rm{S1}=\{\ket{1}[\rm{M}],~\ket{2}[m],~\ket{3}[m]\}, \\
    \rm{S2}=\{\ket{3}[M],~\ket{1}[m],~\ket{2}[m]\}, \\
    \rm{S3}=\{\ket{2}[M],~\ket{1}[m],~\ket{3}[m]\}.
\end{gather}
\label{Eq:Setups}
\end{subequations}
The approximate minority fractions used experimentally in all of the above setups are $f_{\rm{m}} \equiv \frac{N_{\rm{m}}}{N} = 10~\%~(15~\%)$ in setups S1, S3 (S2).
Here, $N_{\rm{m}}$ is the population in the minority components.
Moreover, the intra and interspecies interactions in the different setups are characterized by the $s$-wave scattering lengths $a_{jj'}$, $a_{11}=100.4~a_0$, $a_{22}=100.86~a_0$, $a_{33}=94.57~a_0$, $a_{12}=100.41~a_0$, $a_{13}=98.13~a_0$, $a_{23}=98.98~a_0$, with $a_0$ denoting the Bohr radius [see also Table~\ref{Tab:Interactions} in Methods].
After each spin mixture is produced, we allow the system to evolve in the trap with the potential dimple before using state-selective absorption imaging to measure the spatial density profiles for each component in each of the configurations, see Fig.~\ref{Fig:Spacetimes}.
The potential dimple is used to seed PS formation. 
Even when the dimple is switched off during the time evolution, the identified PS structures remain reproducible, although their formation is delayed.

\textbf{Three-component modeling.}
The stationary and dynamical properties of the three component mixtures are well captured by the following coupled Gross-Pitaevskii equations~\cite{kevrekidis_defocusing_2009,pethick_bose_2008},

\begin{equation}
    i \hbar  \partial_t \Psi_j = \Bigg[ -\frac{\hbar^2 \nabla^2}{2m} + V(\boldsymbol{r})+ \sum_{j'=1}^3  g_{jj'}n_{j'} 
      -i \hbar K^{(j)}_3 n_j^2 \Bigg] \Psi_j,
     \label{Eq:Three_comp_GP}
\end{equation}
where $j=1,2,3$ and $g_{jj'} = 4\pi \hbar^2 a_{jj'}/m$.
Here $\Psi_j=\Psi_j(\boldsymbol{r},t)$ is the macroscopic 3D wavefunction of each component. It is normalized to the respective particle number $\int d\boldsymbol{r} ~ \abs{\Psi_j(\boldsymbol{r},t)}^2 = N_j$, with $\sum_{j=1}^3 N_j = N$ denoting the total particle number. Additionally, $n_j = \abs{\Psi_j}^2$, $m$ is the $^{87}$Rb atomic mass, and $V(\boldsymbol{r}) = \sum_{k=x,y,z} ( m \omega_k^2 k^2/2 ) - V_0 e^{-2x^2/w_x^2-2y^2/w_y^2}$ is the harmonic confinement with the additional central attractive well.  
The attractive dimple is characterized by its height $V_0$ and widths $(w_x,w_y)$, in line with the experimental values stated above.
The $a_{jj'}$ elements  refer to the inter- (intra-) 
component $s$-wave scattering lengths presented above.

To emulate the experimental protocol, all atoms are initialized in the $\ket{1}$ state, confined by $V(\boldsymbol{r})$.
Subsequently, 
the initial atom number is partitioned among three components
realizing the three distinct setups of Eq.~\eqref{Eq:Setups}.
The imaginary contribution models three-body recombination processes of strength $K^{(j)}_3$ for the $j$-th component.
Following the experimental observations we adjust $K^{(j)}_3$, matching the atom losses, which in our setting turn out to be non-negligible only in the $\ket{3}$ state, see the Methods for details. 
The inclusion of losses is crucial for the quantitative agreement with the experimental observations. 
It is relevant to highlight that the presence
of such losses eliminates the potential (of some of the
scenarios considered herein such as, e.g., S1) toward the manifestation of
recurrences that might arise in the Hamiltonian form of the 
system.

\textbf{Reduction scheme.} 
A repulsive, highly particle-imbalanced, two-component gas in the immiscible regime ($a_{12}^2 > a_{11} a_{22}$) can be reduced to an effective single-component attractive one~\cite{Dutton_eff}, a
reduction experimentally leveraged 
for both the Townes soliton~\cite{Bakkali_realization_2021} and the single  PS~\cite{Romero_experimental_2024}.
Here, we {\it generalize} this approach to an arbitrary number of components, see the Methods, 
elucidating the phenomenology observed below.

For the experimental setups described, the presence of one majority and two minority components yields an effective two-component reduction.
To illustrate this, we focus on the time-independent version of Eqs.~\eqref{Eq:Three_comp_GP}, with $K^{(j)}_3=0$, and apply the Thomas-Fermi approximation~\cite{pethick_bose_2008} for the majority component.
Substitution of the majority density back to the equations of the minority species results in an effective two-component system with renormalized scattering lengths (see also the Methods)
\begin{equation}
    a_{m m'}^{(\rm{eff})} = a_{{m} {m}'} - \frac{a_{{m}{M}} a_{{m}'{M}}}{a_{{MM}}}, \quad \forall~ {m},{m}'. \label{Eq:Renorm_interactions} 
\end{equation}
Evidently, Eq.~\eqref{Eq:Renorm_interactions} implies that $a^{(\rm{eff})}_{{m} {m}'}$
can have positive or negative signs even though all intra- and intercomponent scattering lengths between the individual components are positive. 
Hence, different effective attractive two-component setups can be realized dictated by the combination of distinct hyperfine states accommodating the potential formation of vector or multiple RWs.

RW generation in these effective attractive environments is inherently linked to inter- or intra species dominated MI~\cite{goldstein_quasiparticle_1997,kasamatsu_modulation_2006} regulated by $a^{({\rm{eff}})}_{{m} {m}'}$. 
Intra-species MI translates to $a^{({\rm{eff}})}_{{m} \rm{m}}$ being attractive and larger in magnitude than all the other interaction terms.
Inter-species MI is dominant when the following condition 
applies~\cite{kasamatsu_modulation_2006}
\begin{equation}
    |a^{(\rm{eff})}_{{m} {m}'}| > |a^{(\rm{eff})}_{{m} {m}}| \left(  2  \frac{|  a^{({\rm{eff}})}_{{m} {m}} | }{  a^{(\rm{eff})}_{{m}' {m}'}   }  + 1 \right), \quad {m}\neq {m}'.
    \label{Eq:MI_intercomponent_condition}
\end{equation}
 These MI conditions, in turn, determine the existence
of PSs per component, as we now argue.

\begin{figure}[t!]
\centering
\includegraphics[width=1\columnwidth]{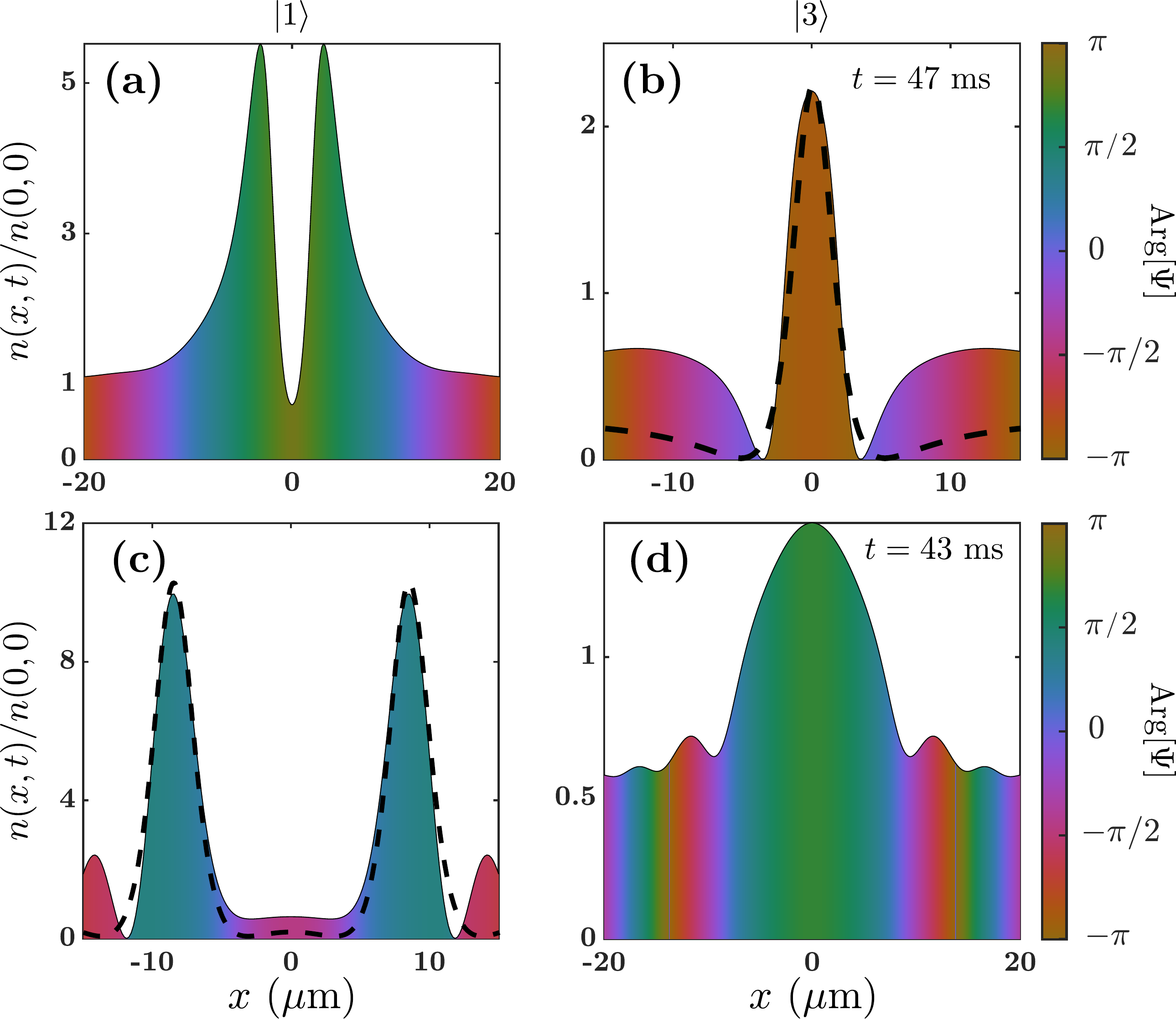}
\caption{\textit{{\bf Twin PS structure.}} Density profiles within the quasi-1D NPGPE of the (a), (c) $\ket{1}$ minority state and the (b), (d) $\ket{3}$ majority component of S2 at selected time-instants (see legends) for (a), (b) $f_{\rm{m}}=15\%$ and (c), (d) $f_{\rm{m}} = 1\%$ population imbalances. 
A twin (single) PS appears in the minority (majority) component for different imbalances captured by the phase (colormap) and the fitted  analytical PS waveform (black dashed lines). Evidently, the twin PS is fully formed for larger imbalances, i.e. smaller $f_{\rm{m}}$.  }
\label{Fig:Phase_profiles}
\end{figure}

\textbf{Formation of vector RWs.}
The mean-field integrated (over the transverse $y$, $z$ directions) density evolution of the individual hyperfine states for the above described three-component setups is shown in Fig.~\ref{Fig:Spacetimes}. 
PS-like nucleation appears in the $\ket{3}$ minority component for S1 (S3) at $t \approx 45 ~\rm{ms}$ ($t \approx 40 ~ \rm{ms}$), traced to the MI of the reduced system.
In fact, both S1 and S3 feature intra-MI in the $\ket{3}$ state [see Supplementary Table I in Supplementary Note 1], which accordingly results in a (bright) PS [Fig.~\ref{Fig:Spacetimes}(a5), (a6), (c5), (c6)].
Absence of intra-/inter-MI in the remaining minority components dictates lack of RWs therein.
In both of these settings, $\ket{2}$ develops a dark (dip)
RW, in a way reminiscent, e.g., of the dark-bright 
RW structures
identified in the integrable Manakov limit~\cite{baronio_vector_2014}.

\begin{figure*}[t!]
\centering
\includegraphics[width=1\textwidth]{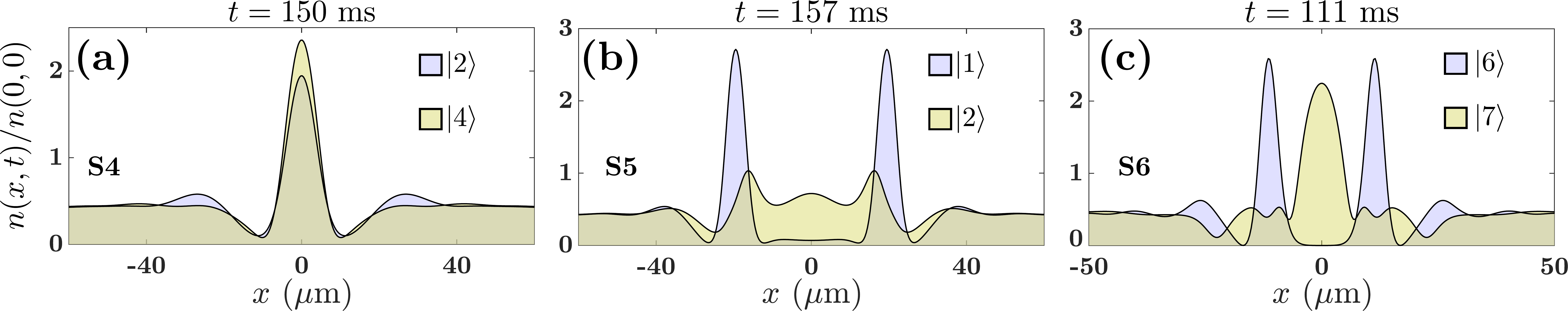}
\caption{\textit{{\bf Zoo of additional PS configurations in three-component BECs.}} Selected density profile snapshots  of different  $^{87}$Rb minority states (see legends) within the NPGPE. 
The three-component settings, $\rm{S4},~\rm{S5}$ and $\rm{S6}$ (all the cases not considered in Fig.~\ref{Fig:Spacetimes}) have fixed particle-imbalance $ f_{\rm{m}} =10 ~ \%$ and demonstrate different (a) vector and (b), (c) twin PS structures. 
} 
\label{Fig:Peregrines_setups}
\end{figure*}

To identify PS formation we confirm the $\pi$ phase jump between center and edge of the ensuing waveform and further fit the numerically generated configuration to the exact, in the integrable limit, analytical solution~\cite{Peregrine1983}, $\Phi_P(x) = \Phi_0 -   4 \Phi_0 \left(1+4 (\frac{x}{L_P})^2\right)^{-1}$. 
$L_P$ is the characteristic length scale of the PS, and $\Phi_0$ the background 
amplitude. 
The PS generation is accompanied by density depletion in the remaining two components. 
This behavior is explained throughout  by relying on the available pairwise immiscibility conditions~\cite{Tommasini,kasamatsu_modulation_2006,kasamatsu_multiple_2004},
emanating from the full three-component system.
Strikingly, immiscibility is dictated by the effective two-component reduction only in the case of extreme imbalances, $f_{\rm{m}} \leq  5~\%$, going  beyond the realm straightforwardly
accessible to our current experimental data.

Specifically, the PS building upon the $\ket{3}$ component is phase immiscible ($a_{3 {M}}^2 \geq a_{33} a_{{M} {M}}$) with the majority $\ket{1}$ ($\ket{2}$) in S1 (S3). 
As such, the latter acquires a dip at the PS location. 
Also, since $a_{{M} {M}} > a_{33}$, the majority extends  
outside the PS configuration.
On the other hand, the PS species is (weakly) immiscible with the  $\ket{2}$ ($\ket{1}$) remaining minority species, and $a_{22} > a_{33}$ ($a_{11} > a_{33}$) implies that in both cases the $\ket{3}$ minority component density accumulates at the trap center.  
Finally, the majority species is miscible with the $\ket{2}$ ($\ket{1}$) minority since $a^2_{{m} {M}} < a_{{m} {m}} a_{{M} {M}}$, with ${M}=1(2)$ ${m}=2(1)$ for S1 and S3 respectively.
Therefore, these two species behave similarly while the majority in S1 (S3) is less (more) spatially extended compared to the remaining minority due to $a_{11} < a_{22}$.
The emerging PS dissolves soon after its creation, giving its place to MI setting in  faster in S3 as compared to S1 due to its associated larger effective attraction (see Eq.~\eqref{Eq:Renorm_interactions} and Table I in the Methods).
Excellent quantitative agreement with the experimental observations can be readily inferred throughout the dynamics, see Figs.~\ref{Fig:Spacetimes}(a1)-(a6), (c1)-(c6). 

Strikingly, in S2 where $\ket{3}$ is the majority species, vector PSs  
develop in all three components in line with the experimental observations.
A splitting of the minority species 
(into two blobs) occurs right after the state preparation 
leading to the gradual formation of a vectorial twin PS state
around $t \approx 35 ~ \rm{ms}$ [Fig.~\ref{Fig:Spacetimes}(b4), (b6)].
Twin PSs appear in both minority components since they suffer intra-MI in addition to the existence of inter-MI [Supplementary Table I in Supplementary Note 1].
Their form is attested through monitoring the phase of the numerically obtained waveform and by fitting to $\Phi_P(x-x_0)\Phi_P(x+x_0)$, with $x_0=3~\rm{\mu m}$ as shown in Fig.~\ref{Fig:Phase_profiles} within the NPGPE approach (see also Supplementary Note 4). 
Since the intraspecies interactions in both minority components are similar, their distributions behave alike. 
In fact, for intermediate population imbalances such as $f_{\rm{m}}=15\%$, a precursor of the twin PS configuration emerges due to significant interactions among the PSs within the same component [Fig.~\ref{Fig:Phase_profiles}(a)], while this configuration fully develops in the extreme population imbalance [Fig.~\ref{Fig:Phase_profiles}(c) and Supplementary Note 2].
It is also found that a $\pi$ phase difference occurs between the PSs building upon different components.
Surprisingly, during the MI stage that is initiated around $t\approx 40 ~ \rm{ms}$, beating dark–dark entities~\cite{Yan_2012} emerge within the filamented densities—previously observed in two-component repulsive condensates~\cite{Hoefer_DD}.

The majority component is immiscible with both minority ones. Since it bears $a_{33}< a_{11}~(a_{33} < a_{22})$, it is pinned between the twin PSs.
It develops a PS structure [Fig.~\ref{Fig:Phase_profiles}(b)] at moderate imbalances, whereas this structure
is significantly deformed
at extreme imbalances [Fig.~\ref{Fig:Phase_profiles}(d)], or in the absence of $K^{(3)}_3$, see Supplementary Note 2.
Importantly, in all three experimental setups the amplitude of the observed PSs are approximately $2.5$ larger compared to the average background. Within the 3D GPE and NPGPE the amplitudes are significantly larger, $4$ times the average background. Such a discrepancy is attributed to the averaging process of independent experimental realizations, which hinders the observation of sharp PS peaks (see also the Methods Section).

\textbf{Heterogeneity of vector PSs.} \label{Sec:More_setups} 
Aside from the three considered setups introduced in Eq.~\eqref{Eq:Setups}, several additional imbalanced three-component settings can be realized, each corresponding to a unique effective attractive environment.
To demonstrate the richness of the system and the generality of our approach, in the following we will briefly discuss three additional systems using states from the $\ket{1,m_F}$ and $\ket{2,m_F}$ hyperfine levels. Namely,
\begin{subequations}
\begin{eqnarray}
    \rm{S4} &= &\{\ket{1}[\rm{M}],~\ket{2}[m],~\ket{4}[m]\}, \\
    \rm{S5}&=&\{\ket{5}[\rm{M}],~\ket{1}[\rm{m}],~\ket{2}[\rm{m}]\}, \\
    \rm{S6}&=&\{\ket{4}[\rm{M}],~\ket{6}[\rm{m}],~\ket{7}[\rm{m}]\}.
\end{eqnarray}
\label{Eq:Setups_additional}
\end{subequations}
Recall that $\rm{M}$ ($\rm{m}$) stands for the majority (minority).
Here, the employed states are $\ket{4} \equiv \ket{2,-2}$, $\ket{5} \equiv \ket{1,1}$, $\ket{6} \equiv \ket{2,-1}$ and $\ket{7} \equiv \ket{2,1}$, while the relevant scattering lengths read $a_{44} = 98.98 ~ a_0$, $a_{14} = 98.98 ~ a_0$, $a_{24} = 97.83~a_0$, $a_{55} = 100.4~a_0$, $a_{15} = 101.32~a_0$, $a_{25}= 100.4~a_0$, $a_{46} = 98.98~a_0$, $a_{47}=92.38~a_0$, $a_{66}=95.68~a_0$, $a_{67}=93.46~a_0$, and $a_{77} = 95.68~a_0$ [see also Table~\ref{Tab:Interactions} in Methods].
We note that the initial conditions for these setups, such as the necessary hyperfine states and mixture ratios, can readily be realized in current experiments. 
However, practical limitations arise from atom losses competing with the envisioned dynamics, particularly in mixtures that include states from the $F=2$ manifold.

The corresponding effective scattering lengths [Eq.~\eqref{Eq:Renorm_interactions}] lead to a distinct interplay of inter- and intra-MI as compared to our earlier findings.
Therefore, for all these settings, a multitude of PS configurations can be generated as predicted by the NPGPE and illustrated in Fig.~\ref{Fig:Peregrines_setups}.
For instance, in S4 the minorities $\ket{2}$ and $\ket{4}$ are miscible forming a PS per component due to the presence of inter-MI [Fig.~\ref{Fig:Peregrines_setups}(a)], since the inequality~\eqref{Eq:MI_intercomponent_condition} is satisfied for $m=2$, and $m'=4$.
They remain at the trap center since they are miscible with the majority $\ket{1}$ state.
This is strongly reminiscent of the ``bright-bright"
vector PS scenario of the attractive Manakov model of~\cite{baronio_2012}. 

Furthermore, considering another hyperfine state combination (S5), where $\ket{1}$ and $\ket{2}$ represent the minorities and $\ket{5}$ the majority, it is possible to generate a twin PS in $\ket{1}$ [Fig.~\ref{Fig:Peregrines_setups}(b)].
The PS origin is traced back to the intra-MI in $\ket{1}$, and the absence of any other MI trait.
The intra-MI character is justified by the fact that $a^{(\rm{eff})}_{11}<0$ and dominates over the other scattering lengths.
Its twin character stems from the immiscibility with the majority. 
The other minority experiences a modulated background with two pronounced peaks, slightly shifted with respect to the PSs of $\ket{1}$. 
Similarly, a twin PS can be generated in the $\ket{6}$ minority state (S6) [see also Fig.~\ref{Fig:Peregrines_setups}(c)] displaying intra-MI (since $a^{(\rm{eff})}_{66}<0$ dominates), while being at the miscibility threshold with the minority $\ket{7}$ (not presenting intra-MI) and immiscible with the majority $\ket{4}$. 
The relevant dynamics further testifying vector RW nucleation for setups S4-S6 is displayed in the Supplementary Note 3.
The partial immiscibility observed among the minorities of S5, S6 fully develops in the extreme imbalance case, and is explained within the reduced model.

{\vspace{0.25cm}\noindent\large\textbf{Conclusions}}
\label{Sec:Conclusions}

We have reported the 
experimental observation of vector PS configurations  arising in repulsive three-component BECs capable of emulating effectively attractive two- and single-component systems.  
To support the analysis we have developed a generic reduction scheme providing the mapping of a repulsive $\mathcal{N}$-component setting to one with 
$\mathcal{N}-k$
minority components, assuming
$k$ majority components, with the
completely general details provided
in the Methods Section.
This substantially expands the results of Refs.~\cite{Dutton_eff,Bakkali_realization_2021,Romero_experimental_2024}, charting
new directions for multi-component
systems.

Exploiting an attractive potential well, we show the controlled and highly reproducible (between
computation and experiment) dynamical formation of a plethora of vector PS structures. 
These include single and twin Peregrines in the components constituting the considered unstable backgrounds suffering intra- or inter-component MI~\cite{kasamatsu_modulation_2006}. 
We showcase ``bright-bright''
and ``dark-bright'' PS states, not only 
materializing, but also generalizing
states that were earlier predicted
in similar settings~\cite{baronio_2012,baronio_vector_2014}.
Our experimental observations are in quantitative agreement with 3D mean-field simulations but also with quasi-1D NPGPE ones. 
Leveraging the latter and expanding upon the experimental flexibility of three-component setups corroborated by our effective model reduction we reveal a multitude of vector PSs.

Our experiments provide a stepping stone for realizing exotic vector or higher-order RWs~\cite{Chabchoub_water1}, probing their collisional and interaction features in a controllable way using cold atoms. 
We outline some of the associated 
emerging opportunities below.
A further
theoretical and computational analysis of the effective models 
that we bring forth herein and of their 
palette of possible solutions
would advance our understanding in a highly non-trivial manner. 
Another interesting direction to pursue is the nucleation of RW solutions that are periodic either in space or time, such as the Akmediev~\cite{akhmediev2009waves} or Kuznetzov-Ma~\cite{ma1979perturbed} 
in  multi-component settings, which
indeed
awaits for experimental observation in ultracold atoms (among other disciplines). 
The consideration of a larger number of 
effective components, e.g., in the realm of a spin-2
BEC~\cite{stamper2013spinor} is also a natural 
possibility for future studies.
Finally, exploring the role of genuine quantum effects in RW formation, which can be achieved by utilizing perturbative treatments to the mean-field energy functional or sophisticated many-body computations~\cite{mistakidis2023few}, is of considerable  interest, with the
relevant direction being wide open 
in the multi-component realm.

{\vspace{0.25cm}\noindent\large\textbf{Methods}}  \label{Sec:Methods}

\textbf{Experimental cross-sections and losses.}

\begin{figure}[h]
    \centering
    \includegraphics[width=1\columnwidth]{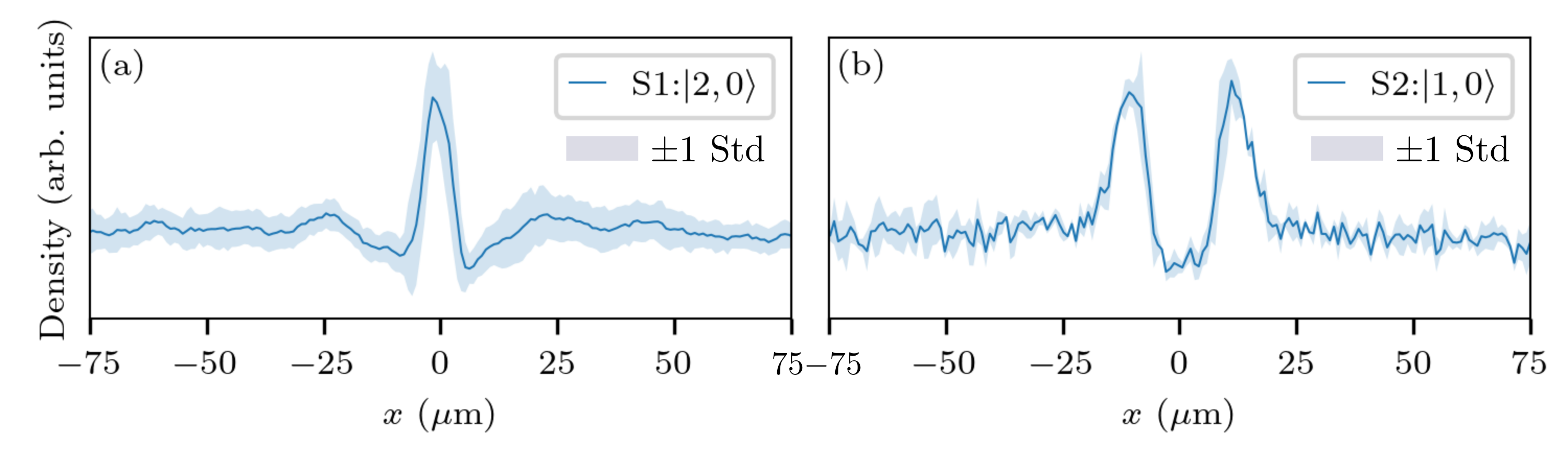}
    \caption{\textbf{Cross-sections of PSs.} Averaged integrated cross sections over 15 experimental realizations of (a) Setup S1 and (b) Setup S2 at evolution times of 50 ms and 35 ms, respectively. Shading represents one standard deviation (Std) of variation in the density.}
    \label{fig:cross_sections}
\end{figure}

Representative experimental cross-sections capturing the single and twin PS nucleation in the minority components $\ket{3}$ and $\ket{1}$ of the S1 and S2 settings  discussed in the main text are shown in Fig.~\ref{fig:cross_sections} along with their standard deviations (see shaded areas) stemming from 15 different experimental realizations.
The reproducibility of the observed Peregrine-like structures can be deduced from the relatively small standard deviations, but it is important to note that the sharpness of the Peregrine features in the experimental cross sections is blunted due to the averaging over experimental realizations.

\begin{figure}[h]
    \includegraphics[width=1\columnwidth]{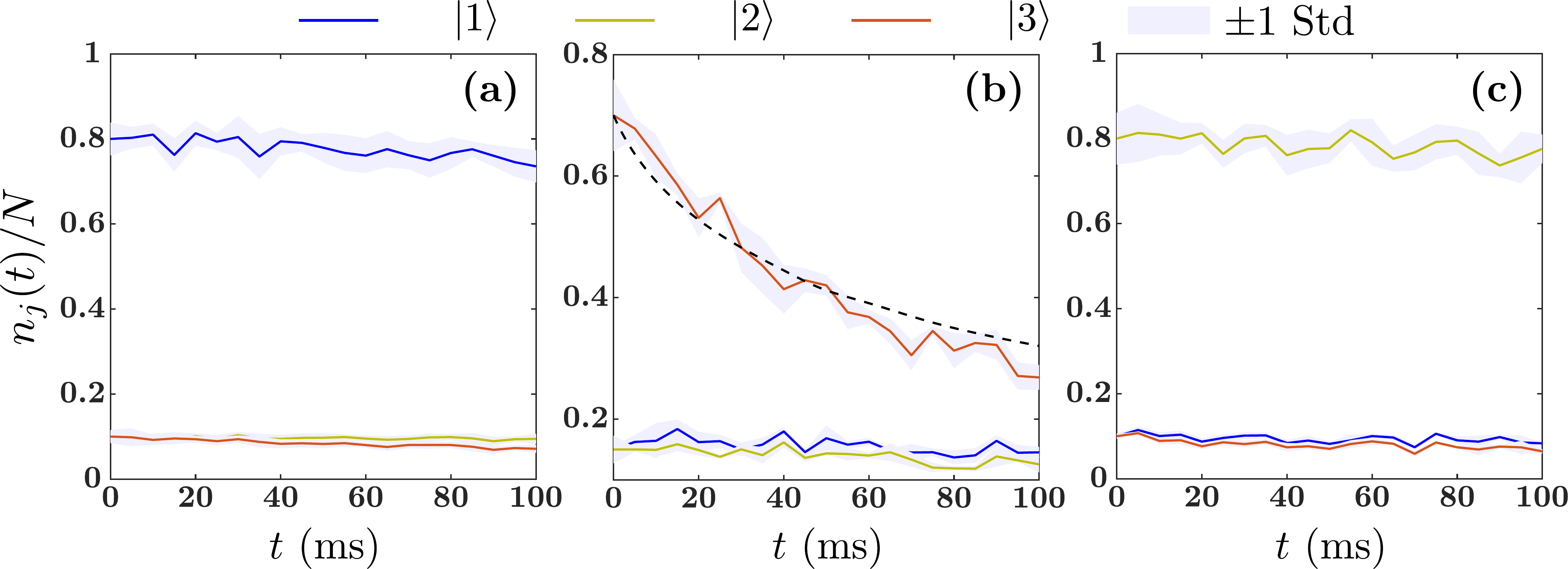}
    \caption{\textbf{Atom loss dynamics.} (a)-(c) Experimentally extracted population fraction dynamics, $n_j(t)/N$ with $j=1,2,3$, of the individual hyperfine states (see legends) participating in the three different Setups S1-S3 (from left to right), discussed in the main text. The shaded regions indicate one standard deviation (Std) from the mean population fractions (solid lines). It can be seen that the $\ket{3} $ hyperfine state features the most prominent losses. The dashed line in panel (b) represents the fit characterizing three-body losses used in the simulations.}
    \label{Fig:Losses}
\end{figure}

\begin{table*}[t]
    \centering
    \begin{tabular}{|c|c|c|c||c|c|c|c|}
    \hline 
        \textbf{S1}  & $\ket{1,-1}$~[M] & $\ket{1,0}$~($a^{(\rm{eff})}$) & $\ket{2,0}$  &  \textbf{S2}  &  $\ket{2,0}$~[M] & $\ket{1,-1}$ & $\ket{1,0}$   \\ \hline
        $\ket{1,-1}$~[M]  & 100.4 & 100.41 & 98.13 & $\ket{2,0}$~[M]  & 94.57 & 98.13 & 98.98 \\   \hline
        $\ket{1,0}$ & - & 100.86 (0.44) & 98.98 (0.84) & $\ket{1,-1}$ & - & 100.4 (-1.42) & 100.41 (-2.30) \\  \hline
        $\ket{2,0}$ &  - & -  & 94.57 (-1.34) & $\ket{1,0}$ & - & - & 100.86 (-2.74)  \\  \hline
         \hline
        \textbf{S3}  & $\ket{1,0}$~[M] & $\ket{1,-1}$ & $\ket{2,0}$ &  \textbf{S4}  & $\ket{1,-1}$~[M] & $\ket{1,0}$ & $\ket{2,-2}$  \\ \hline
        $\ket{1,0}$~[M]  & 100.86 & 100.41 & 98.98 & $\ket{1,-1}$~[M]  & 100.4 &  100.41  &  98.98  \\  \hline
        $\ket{1,-1}$ & - & 100.4 (0.44) & 98.13 (-0.41) & $\ket{1,0}$ & - & 100.86 (0.44) & 97.85 (-1.14)  \\  \hline
        $\ket{2,0}$ & - & - & 94.57 (-2.57) & $\ket{2,-2}$ & - & - & 98.98 (1.4) \\  \hline
         \hline
        \textbf{S5}  & $\ket{1,1}$~[M] & $\ket{1,-1}$ & $\ket{1,0}$ & \textbf{S6}  & $\ket{2,-2}$~[M] & $\ket{2,-1}$ & $\ket{2,1}$  \\ \hline
        $\ket{1,1}$~[M]  & 100.4 &  101.32 &  100.40 & $\ket{2,-2}$~[M]  & 98.98 &  98.98 & 92.38 \\  \hline
        $\ket{1,-1}$ & - & 100.4 (-1.85) & 100.41 (-0.91) & $\ket{2,-1}$ & - & 95.68 (-3.3) &  93.46 (1.1) \\  \hline
        $\ket{1,0}$ & - & - & 100.86 (0.46) & $\ket{2,1}$ & - & - & 95.68 (9.46) \\  \hline
    \end{tabular}
    \caption{$s$-wave scattering lengths $a_{ij}$ (in Bohr radii) between the different hyperfine levels of six distinct three-component settings denoted by (\textbf{S1}-\textbf{S6}). The majority component in all cases is indicated by [M]. The values in the parentheses correspond to the effective scattering lengths of the underlying effective two-component system consisting of only the minorities (see also Eq.~(2)). Due to the $a_{ij} = a_{ji}$ symmetry some values in the table (marked by -) are omitted for brevity.}
    \label{Tab:Interactions}
\end{table*}

Three-body recombination~\cite{Greene_universal_2017} is the prominent atom loss mechanism, due to the high density of the vector PSs.
Indeed, the inclusion of this loss process within our mean-field (3D and NPGPE) simulations results in better quantitative agreement with the experimental data. 
We have verified this statement by comparing our simulations in the absence of losses or by solely considering two-body losses. 
To estimate the appropriate loss coefficients $K^{(j)}_3$ (see also Eq.~\eqref{Eq:Three_comp_GP}, we employ the {\it approximate} rate equations~\cite{Greene_universal_2017}, $\dot{n}_j(t) = -K^{(j)}_3 n_j^3(t)$.
The latter admits the formal solution, $n_j(t) = \frac{n_{j,0}}{\sqrt{1+2K^{(j)}_3 t n^2_{j,0}}}$, where $n_{j,0}$ is the initial (uniform) density of the $j$-th component.
$K^{(j)}_3$ can be estimated from the population fraction of each component at the end of the evolution, $n_j(t = 100 ~ \textrm{ms}) = \lambda_j n_{j,0}$, i.e. $K^{(j)}_3 = \frac{1-\lambda_{j,0}^2}{200 n_{j,0}^2 \lambda_j^2}~\rm{kHz}$.
The atom loss resulting from this matching process is in very good agreement with the experimental loss data of the $\ket{3}$ hyperfine level, as can be seen in Fig.~\ref{Fig:Losses}(b), which provides an overview of the atom losses of the individual hyperfine levels for Setups S1-S3 discussed in the main text.
It becomes apparent that significant losses take place in $\ket{3}$, which become more prominent when this state represents the majority component as in S2 [Fig.~\ref{Fig:Losses}(b)].

\textbf{General reduction of an $\mathcal{N}$-component setting.} 

Here, we outline the scheme for reducing a general $\mathcal{N}$ component system to an effective $\mathcal{N}-k$, where $k$ is the number of majority components, $1 \leqslant k \leqslant \mathcal{N}-1$. 
The starting point is the $\mathcal{N}$ coupled time-independent Gross-Pitaevskii  equations (GPEs),

\begin{equation}
    \mu_i \Psi_i = -\frac{\hbar^2}{2m} \nabla^2 \Psi_i +V(\boldsymbol{r}) \Psi_i +\sum_{j=1}^{\mathcal{N}} g_{ij} n_j \Psi_i, ~ i=1,\ldots,\mathcal{N},
    \label{Eq:General_GPs}
\end{equation}
where $n_j = \abs{\Psi_j}^2$, $g_{ij} = 4\pi \hbar^2 a_{ij} / m$, and $\mu_i$ is the chemical potential of the $i$-th component.

In our notation, we order the component indices, so that the first $k$ correspond to the majority components, and the rest to the minority components.
Employing the Thomas-Fermi approximation
to the majority components, the equations for the latter are recast as follows
\begin{equation}
\mu_i = V(\boldsymbol{r}) + \sum_{j=1}^{\mathcal{N}} g_{ij} n_j, ~ i=1,\ldots,k.
\label{Eq:Majority_eqs}
\end{equation}
Having this at hand it is possible to determine the densities of the majority components.
To do so, we exploit a more compact matrix notation, by considering the symmetric interaction strength matrix $\boldsymbol{g}$, whose elements read $\boldsymbol{g}_{ij}=g_{ij}, ~ i,j=1,\ldots,\mathcal{N}$.
Moreover, we introduce the projections to the subspaces of majority and minority components, which can be respectively expressed as 
\begin{subequations}
\begin{gather}
     \boldsymbol{M} = \textbf{diag} \left[ \underbrace{1,\ldots,1}_k, \underbrace{0,\ldots,0}_{\mathcal{N}-k}  \right], \\
   \boldsymbol{m} = \textbf{diag} \left[ \underbrace{0,\ldots,0}_k, \underbrace{1,\ldots,1}_{\mathcal{N}-k}  \right] \label{Eq:Projections},
\end{gather}
\end{subequations}
where $\textbf{diag}$ denotes a diagonal matrix.
Since these are orthogonal projections, they satisfy $\boldsymbol{M}^2=\boldsymbol{M}$, $\boldsymbol{m}^2 = \boldsymbol{m}$, and $\boldsymbol{M} \cdot\boldsymbol{m}=\boldsymbol{0}$, while $\cdot$ designates matrix multiplication. Equipped with these projections, we can now define the interaction strengths within each subspace, as well as the couplings between majority and minority components,

\begin{subequations}
\begin{gather}
    \boldsymbol{g}_{\rm{M}}  \equiv \boldsymbol{M} \cdot \boldsymbol{g} \cdot \boldsymbol{M}, \quad 
    \boldsymbol{g}_{\rm{m}} \equiv \boldsymbol{m} \cdot \boldsymbol{g} \cdot \boldsymbol{m}, \label{Eq:Intra_interactions} \\
    \boldsymbol{g}_{\rm{Mm}}  \equiv \boldsymbol{M} \cdot \boldsymbol{g} \cdot \boldsymbol{m}, \quad
    \boldsymbol{g}_{\rm{mM}}  \equiv \boldsymbol{m} \cdot \boldsymbol{g} \cdot \boldsymbol{M}. \label{Eq:Mixed_interactions}
\end{gather}
\end{subequations}
Similarly, one can define the densities and chemical potentials pertaining only to one subspace,
\begin{subequations}
\begin{gather}
    \boldsymbol{n}_{\rm{M}} \equiv \boldsymbol{M} \cdot \boldsymbol{n}, \quad \boldsymbol{n}_{\rm{m}} \equiv \boldsymbol{m} \cdot \boldsymbol{n}, \label{Eq:Density_defintions} \\
    \boldsymbol{\mu}_{\rm{M}} \equiv \boldsymbol{M} \cdot \boldsymbol{\mu}, \quad \boldsymbol{\mu}_{\rm{m}} \equiv \boldsymbol{m} \cdot \boldsymbol{\mu}, \label{Eq:Chemicals_defintions}
\end{gather}
\end{subequations}
with $ \boldsymbol{n} = \left[  n_1, \ldots, n_{\mathcal{N}}   \right]^{\intercal}$, and $\boldsymbol{\mu} = \left[  \mu_1, \ldots, \mu_{\mathcal{N}} \right]^{\intercal}$.

Using the above, Eq.~\eqref{Eq:Majority_eqs} can be re-formulated in terms of compact matrix notation. Explicitly,

\begin{gather}
    \mu_i = V(\boldsymbol{r}) + \sum_{j=1}^k g_{ij} n_j +\sum_{j=k+1}^{\mathcal{N}} g_{ij} n_j, ~ i=1,\ldots,k  \nonumber \\  
    \Longrightarrow
    \boldsymbol{\mu}_{\rm{M}} = V(\boldsymbol{r}) \boldsymbol{1}_{\rm{M}} + \boldsymbol{g}_{\rm{M}} \cdot \boldsymbol{n}_{\rm{M}} +\boldsymbol{g}_{\rm{Mm}} \cdot \boldsymbol{n}_{\rm{m}} 
    \label{Eq:Majority_matrix_notation}.
\end{gather}
In the above, the projection of the unity vector to the majority subspace has been employed, $\boldsymbol{1}_M = \boldsymbol{M} \cdot \boldsymbol{1}$.
Eq.~\eqref{Eq:Majority_matrix_notation} is now inverted to obtain the densities of the majority components,
\begin{equation}
    \boldsymbol{n}_{\rm{M}} = \boldsymbol{g}_{\rm{M}}^{-1} \cdot \Big[  \boldsymbol{\mu}_{\rm{M}} -V(\boldsymbol{r}) \boldsymbol{1}_{\rm{M}} -\boldsymbol{g}_{\rm{Mm}} \cdot \boldsymbol{n}_{\rm{m}} \Big].
    \label{Eq:Majority_inversion}
\end{equation}

The first step in determining an effective $\mathcal{N}-k$ setup is to cast the $\mathcal{N}-k$ equations in  matrix formalism,
\begin{gather}
    \mu_i \Psi_i = -\frac{\hbar^2}{2m} \nabla^2 \Psi_i + V(\boldsymbol{r}) \Psi_i +\sum_{j=1}^k g_{ij} n_j \Psi_i   \nonumber \\
    + \sum_{j=k+1}^{\mathcal{N}} g_{ij} n_j \Psi_i, ~ i=k+1,\ldots,\mathcal{N}  \Longrightarrow  \nonumber \\
    \boldsymbol{\Psi}_{\rm{m}} \odot \boldsymbol{\mu}_{\rm{m}} = -\frac{\hbar^2}{2m} \nabla^2 \boldsymbol{\Psi}_{\rm{m}} + V(\boldsymbol{r}) \boldsymbol{\Psi}_{\rm{m}} + \boldsymbol{\Psi}_{\rm{m}} \odot \left[ \boldsymbol{g}_{\rm{m}} \cdot \boldsymbol{n}_{\rm{m}} \right] \nonumber \\ + \boldsymbol{\Psi}_{\rm{m}} \odot \left[ \boldsymbol{g}_{\rm{mM}} \cdot \boldsymbol{n}_{\rm{M}}  \right],
    \label{Eq:Minority_matrix}
\end{gather}
where $\odot$ denotes the Hadamard product, while $\boldsymbol{\Psi}_{\rm{m}} = \boldsymbol{m} \cdot \boldsymbol{\Psi}$, and $\boldsymbol{\Psi} = \left[ \Psi_1,\ldots,\Psi_{\mathcal{N}}  \right]^{\intercal}$.
Subsequently, the expression for the majority species  densities is substituted from Eq.~\eqref{Eq:Majority_inversion}. Rearranging terms, one arrives at the effective equations for the minority components,
\begin{gather}
    \boldsymbol{\Psi}_{\rm{m}} \odot \left[ \boldsymbol{\mu}_{\rm{m}} - \boldsymbol{g}_{\rm{mM}} \cdot \boldsymbol{g}_{M}^{-1} \cdot \boldsymbol{\mu}_{\rm{M}} \right] = 
    -\frac{\hbar^2}{2m} \nabla^2 \boldsymbol{\Psi}_{\rm{m}} \nonumber \\ + \boldsymbol{\Psi}_{\rm{m}} \odot \left[ \boldsymbol{1}_{\rm{m}} -  \boldsymbol{g}_{\rm{mM}} \cdot \boldsymbol{g}_{\rm{M}}^{-1} \cdot \boldsymbol{1}_{\rm{M}} \right] V(\boldsymbol{r}) \nonumber \\ + \boldsymbol{\Psi}_{\rm{m}} \odot \Big\{ \left[  \boldsymbol{g}_{\rm{m}} - \boldsymbol{g}_{\rm{mM}} \cdot \boldsymbol{g}_{\rm{M}}^{-1} \cdot \boldsymbol{g}_{\rm{Mm}}  \right] \cdot \boldsymbol{n}_{\rm{m}} \Big \}.
    \label{Eq:Minority_reduced}
\end{gather}

Since the majority component is integrated out, the chemical and trapping potentials, as well as the interactions become renormalized according to
\begin{subequations}
\begin{gather}
    \boldsymbol{\mu}_{\rm{m}}^{(\rm{eff})} = \boldsymbol{\mu}_{\rm{m}} - \boldsymbol{g}_{\rm{mM}} \cdot \boldsymbol{g}_{\rm{M}}^{-1} \cdot \boldsymbol{\mu}_{\rm{M}}, \\
    \boldsymbol{V}(\boldsymbol{r})^{(\rm{eff})} =   V(\boldsymbol{r}) \left[  \boldsymbol{1}_{\rm{m}}-   \boldsymbol{g}_{\rm{mM}} \cdot \boldsymbol{g}_{\rm{M}}^{-1} \cdot \boldsymbol{1}_{\rm{M}} \right], \\
    \boldsymbol{g}_{\rm{m}}^{(\rm{eff})} =  \boldsymbol{g}_{\rm{m}} - \boldsymbol{g}_{\rm{mM}} \cdot \boldsymbol{g}_{\rm{M}}^{-1} \cdot \boldsymbol{g}_{\rm{Mm}}.
    \label{Eq:Renormalizations}
\end{gather}
\end{subequations}
Note that the effective potential is in principle different for the minority components.
Below, we list a few explicit expressions of the effective interactions for different setups,
\begin{subequations}
\begin{gather}
    \mathcal{N}=2, \quad k=1 \Longrightarrow g^{(\rm{eff})}_{\rm{m m}} = g_{\rm{mm}} - \frac{g^2_{\rm{mM}}}{g_{\rm{MM}}}, \label{Eq:Two_one_reduction} \\
    \mathcal{N}=3, \quad k=1 \Longrightarrow g^{(\rm{eff})}_{\rm{m m'}} = g_{\rm{m m'}} - \frac{g_{\rm{mM}} g_{\rm{m' M}}}{g_{\rm{MM}}}, \label{Eq:Three_two_reduction} \\
    \mathcal{N}=3, \quad k=2 \Longrightarrow g^{(\rm{eff})}_{\rm{m m}} = \Bigg[    g^2_{\rm{M m}} g_{\rm{M' M'}}  +g^2_{\rm{M M'}} g_{\rm{m m}}    \nonumber \\   +g_{\rm{M M}} (g^2_{\rm{M' m}}-g_{\rm{M' M'}}  g_{\rm{m m}})  - 2 g_{\rm{M M'}} g_{\rm{M m}} g_{\rm{M' m}} \Bigg]    \nonumber \\
    \times  \Bigg[ \frac{1}{g^2_{\rm{M M'}} - g_{\rm{MM}} g_{\rm{M' M'}}  }  \Bigg ]. \label{Eq:Three_one_reduction}
\end{gather}
\end{subequations}
The $\rm{M}$ ($\rm{m}$) indices pertain to the majority (minority) components.
The reduction of a binary immiscible mixture to an effectively attractive single component setting was first utilized to probe Townes solitons in Refs.~\cite{Bakkali_realization_2021,bakkali_townes_2022}, employing the same effective interaction as in Eq.~\eqref{Eq:Two_one_reduction}.
However, the reduction of a three-component setting with one majority species [Eq.~\eqref{Eq:Three_two_reduction}] employed in this work leads to a far richer phenomenology of effective two-component setups with tunable interactions in both  magnitude and sign. 
This is exemplarily demonstrated for six different experimentally relevant combinations of hyperfine levels in Table~\ref{Tab:Interactions}, whose  effective scattering lengths are provided inside the parentheses.

{\vspace{0.25cm}\noindent\large\textbf{Data Availability}}
The data associated with this work are available from the corresponding author upon request.

{\vspace{0.25cm}\noindent\large\textbf{Code Availability}}
The codes associated with this work are not publicly deposited since they are under development for use in future publications, but they are available upon request from the corresponding author.

{\vspace{0.25cm}\noindent\large\textbf{Author Competing Interests}}
The authors declare no competing interests.

{\vspace{0.25cm}\noindent\large\textbf{Author Contributions}}
The experimental data were taken and analyzed by P.E. and S.M.. G.C.K., S.I.M. and G.A.B. carried out the simulations of the 3D GPE and NPGPE, and analyzed the resulting data. G.A.B. implemented the general reduction scheme, under the guidance of P.G.K., G.C.K. and S.I.M.. All authors contributed to the preparation of the manuscript.

{\vspace{0.25cm}\noindent\large\textbf{Acknowledgements}}

P.G.K. acknowledges support from the Okinawa Institute of Science and
Technology (OIST) via the Theoretical Sciences Visiting Program (TSVP). 

{\vspace{0.25cm}\noindent\large\textbf{Funding}}

This material is partially based upon work supported by the U.S. National Science Foundation under the awards PHY-2110030, DMS-2204702 and PHY-2408988 (PGK). P. E. acknowledges support from the NSF through Grant No. PHY-2207588
and from a Boeing Endowed Professorship
at WSU. 
S.I.M. acknowledges support by the Army Research Office (ARO) under Award number: W911NF-26-1-A043. 
This work was also 
supported by a grant from the Simons Foundation
[SFI-MPS-SFM-00011048, P.G.K].

\putbib[Three_comps]

\end{bibunit}

\clearpage

\begin{bibunit}[apsrev4-1]


\onecolumngrid
\setcounter{equation}{0}
\setcounter{figure}{0}
\setcounter{section}{0}
\makeatletter
\renewcommand{\figurename}{Supplementary FIG.}
\renewcommand{\tablename}{Supplementary TABLE}
\renewcommand{\theequation}{S\arabic{equation}}
\renewcommand{\bibnumfmt}[1]{[C#1]}
\renewcommand{\citenumfont}[1]{C#1}
\renewcommand{\thesection}{\arabic{section}}
\setcounter{page}1
\def\thepage{S\arabic{page}}

\begin{center}
	{\Large\bfseries Supplementary Material: Observation of vector rogue waves in repulsive three-component atomic mixtures \\ 
 }
\end{center}

\section{Supplementary Note 1: Intra- and Inter-component MI}\label{sec:interMI}

The interplay between the attractive inter- and intra-species interactions is related to the presence of intra- and inter-species dominated modulational instability (MI) for the effective  two-component model~\cite{kasamatsu_modulation_2006}.
Intra-species MI occurs when $g^{(\rm{eff})}_{\rm{mm}}$ is attractive, 
whereas the inter-species MI 
dominates the dynamics in the case of  
$|g^{(\rm{eff})}_{\rm{m} \rm{m}'}| > |g^{(\rm{eff})}_{\rm{m} \rm{m}}| \left(  2  \frac{|  g^{(\rm{eff})}_{\rm{m} \rm{m}} | }{  g^{(\rm{eff})}_{\rm{m}' \rm{m}'}   }  + 1 \right)$, $\rm{m} \neq \rm{m'}$~\cite{kasamatsu_modulation_2006}.
Both are determined by the dispersion relation 
of the reduced two-component analogue~\cite{kasamatsu_modulation_2006}
\begin{equation}
\omega^2_{\pm}(k_{\rm{m}},k_{\rm{m'}}) = \frac{1}{2}  \left[ \omega^2_{\rm{m}} + \omega^2_{\rm{m'}}  \pm \sqrt{(\omega^2_{\rm{m}} + \omega^2_{\rm{m'}})^2 + 4(Q^2-\omega^2_{\rm{m}} \omega^2_{\rm{m'}})}   \right],
\label{Eq:Dispersion_relation}
\end{equation}
where $\omega^2_{\rm{m}} = \frac{\hbar^2 k^2_{\rm{m}}}{2m} \left(  \frac{\hbar^2 k^2_{\rm{m}}}{2m} + 2 g^{(\rm{eff})}_{\rm{m m}} n_0   \right)$, and $Q = \frac{\hbar^2 g^{(\rm{eff})}_{\rm{m m'}}}{m} k_{\rm{m}} k_{\rm{m'}} n_0$. Here, $n_0$ denotes the initial local density at the trap center of both minority components.
In order to derive such a dispersion relation, perturbations on top of the initial profile assume the form of plane waves.
Note that this 
dispersion relation 
applies only to homogeneous binary mixtures.
The $\omega_+$ branch is associated with the speed of sound of the effective binary mixture~\cite{mossman2024nonlinear}, whereas finite imaginary $\omega_-$ indicates the presence of MI.
The intra-species MI character of setup S1 is depicted in Supplementary Fig.~\ref{Fig:Dispersion_relations}(a) by a finite $\Im[\omega_-]$ ribbon, stretching along a single momentum axis.
In contrast, setup S2 exhibits a dual MI character [see also Supplementary Table~\ref{Tab:MI_characteristics}], and therefore $\Im[\omega_-]$ displays two finite overlapping regions.
In this scenario, the maximum growth rate, $\max\{ \Im[\omega_-] \}$, is larger than in the case of setup S1, implying that the MI onset is expected to be faster in setup S2 which can be also inferred from Fig.~1 of the main text. 

\begin{figure}[h!]
\centering
\includegraphics[width=0.7\textwidth]{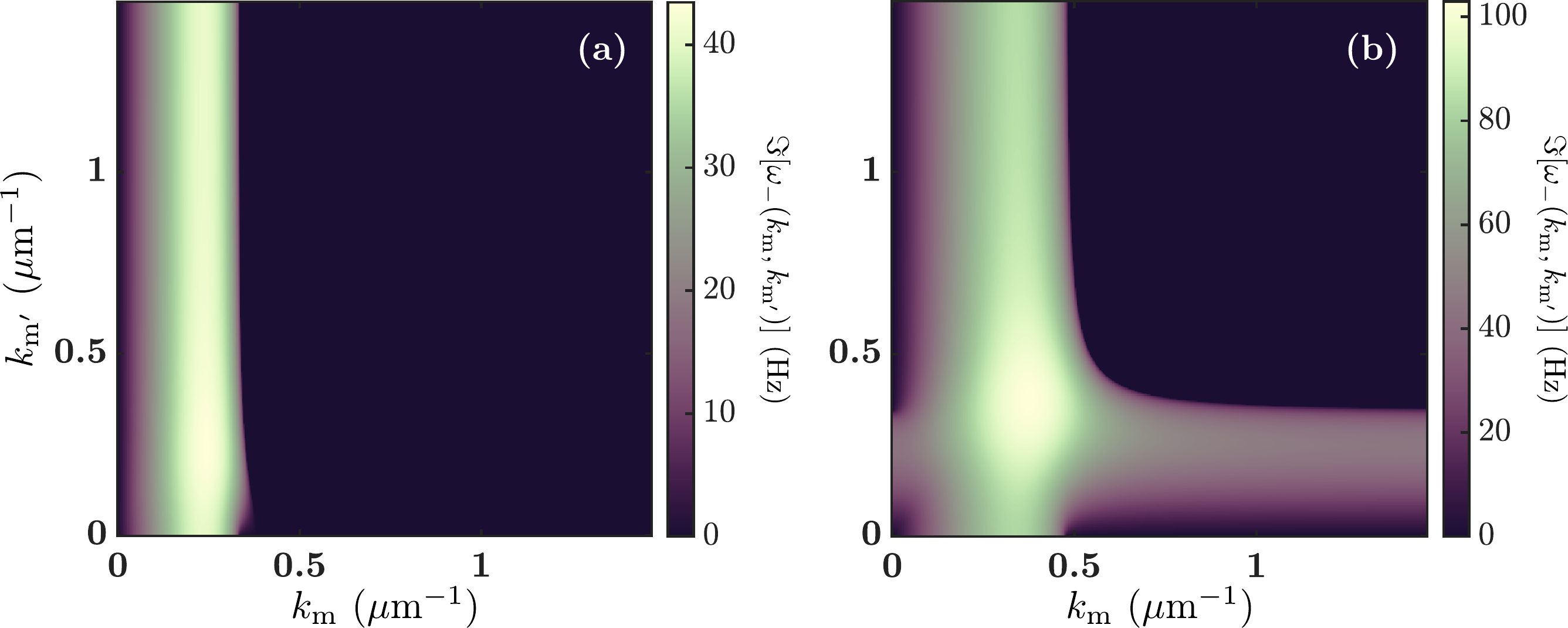}
\caption{\textbf{Inter- and intra-species MI manifestation in the dispersion relations of the reduced two-component model}. MI instability materializes by the finite imaginary values of the $\omega_-(k_{\rm{m}},k_{\rm{m'}})$ dispersion relation branch of the reduced two-component system [see Eq.~\eqref{Eq:Dispersion_relation}]. The intra-species MI character in setup S1 [panel (a)] is evidenced by the elongated $\Im[\omega_-]$ finite ribbon along one momentum axis, whereas the dual inter- and intra-species MI behavior of setup S2 [panel (b)] is visualized by the two overlapping $\Im[\omega_-]$ ribbons.}
\label{Fig:Dispersion_relations}
\end{figure}

MI is generally anticipated to be a  
necessary condition for PS generation, and thus the presence or absence of inter- and intra-species MI provides a promising indicator for PS generation for the 
minority component to which it applies.
The MI behavior of each minority component deduced from Eq.~(3) in the main text for the effective interactions of the reduced two-component system is summarized in Supplementary Table~\ref{Tab:MI_characteristics} for the six distinct experimentally relevant three-component setups. 
At least one PS builds upon each component only in Setups S2 and S4 featuring 
dominant inter-species MI.
All the other settings are subject to intra-MI in one of the minority components, and hence at least one PS appears in the respective minority component. 
Finally, the potential pairwise miscibility of the minority species as dictated by the interaction strengths of the three-component system is shown in the fifth and sixth columns of Suplementary Table~\ref{Tab:MI_characteristics}.

\begin{figure}[h]
    \centering
    \includegraphics{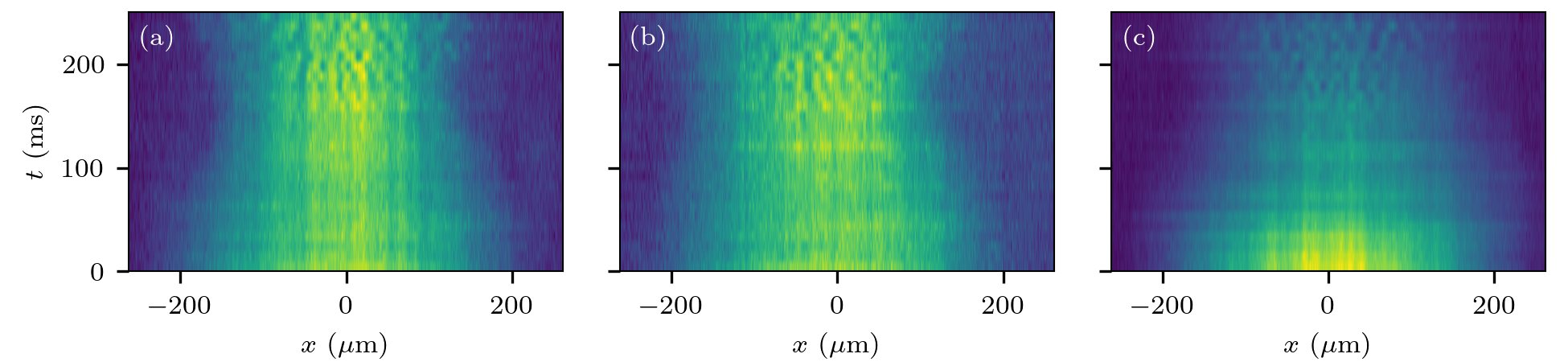}
    \caption{Experimental spacetime dynamics corresponding to Setup S2 from the main text, in the absence of the attractive well, with each spin state imaged independently in (a) $\ket{1} \equiv \ket{1,-1}$, (b) $\ket{2} \equiv \ket{1,0}$, and (c) $\ket{3} \equiv \ket{2,0}$. 
    Each row of the spacetime diagram is a linear cross section of the density averaged over 6 independent experimental realizations, with the image brightness normalized to the atomic density at $t=0$ for each component.
    Fluctuations attributed to MI set in at approximately $t=150$~ms for all three components, in accordance with the effective model prediction for intra- and inter-species MI [see Supplementary Table~\ref{Tab:MI_characteristics}].}
    \label{fig:S2_nobarrier}
\end{figure}

To confirm the presence of intra- and inter-component MI we have performed additional experiments in the absence of the optical barrier used in the main text to seed the PS. 
In what follows, we showcase the 
emergence of MI in a spin mixture with one majority and two minority
components in the case of setup S2, with the minority components making up 15\% each.
Supplementary Fig.~\ref{fig:S2_nobarrier} shows density distributions over time, each row being averaged over six independent experimental realizations. The density fluctuations associated with MI setting in are observed around $150~$ms, beginning first where the density is the highest in the center of the condensate. 

\begin{table}[t!]
\centering
\renewcommand{\arraystretch}{1.2} 
\setlength{\tabcolsep}{8pt} 
\begin{tabular}{|c||c|c|c|c|c|c|c|c|}
\hline
\ding{93}  & Inter MI  & Intra MI ($1^{\rm{st}}$) & Intra MI ($2^{\rm{nd}}$) & Miscible & Immiscible & Peregrine~(component) & Effective    \\
\hline
\textbf{S1}  &  \ding{55}   &  \ding{55} & \ding{51}   & \ding{55}  & \ding{51}  &  $0(1^{\rm{st}})$,~$1(2^{\rm{nd}})$ &  Immiscible  \\  \hline
\textbf{S2}  & \ding{51}  & \ding{51} & \ding{51} & \ding{51}  & \ding{55} &  $2(1^{\rm{st}})$,~$2(2^{\rm{nd}})$ & Immiscible  \\  \hline
\textbf{S3} & \ding{55}  & \ding{55} &  \ding{51} & \ding{55}  & \ding{51}  & $0(1^{\rm{st}})$,~$1(2^{\rm{nd}})$  & Immiscible \\ \hline
\textbf{S4}  & \ding{51} & \ding{55} & \ding{55} & \ding{51} & \ding{55}  &  $1(1^{\rm{st}})$,~$1(2^{\rm{nd}})$  & Immiscible  \\  \hline
\textbf{S5} & \ding{55} & \ding{51} & \ding{55}  & \ding{51} & \ding{55}  & $2(1^{\rm{st}})$,~$0(2^{\rm{nd}})$  & Immiscible  \\  \hline
\textbf{S6} & \ding{55}  & \ding{51} &  \ding{55} & \ding{51} & \ding{55} & $2(1^{\rm{st}})$,~$0(2^{\rm{nd}})$ & Immiscible \\
\hline
\end{tabular}
\caption{Characteristics of the minority components for the six different setups presented in the main text. 
The intra- and inter-species MI [see also Supplementary Fig.~\ref{Fig:Dispersion_relations}] stems from the reduced two-component model [Eq.~(2) in the main text], while the pairwise miscibility among the minority species stems from the interaction strengths [Table I in the main text] of the genuine three-component model. 
The number of PSs per minority component is also reported with the presence (absence) of checkmarks (crosses). Generation of PSs is ensured by the presence of inter- and/or intra-component MI. The miscibility dictates the spatial distribution of the components and PS number. For extreme imbalances $f_{\rm{m}} < 5\%$ the effective two-component reduction dictates immiscibility for all setups.}
\label{Tab:MI_characteristics}
\end{table}

\begin{figure}
    \centering
    \includegraphics[width=0.8\textwidth]{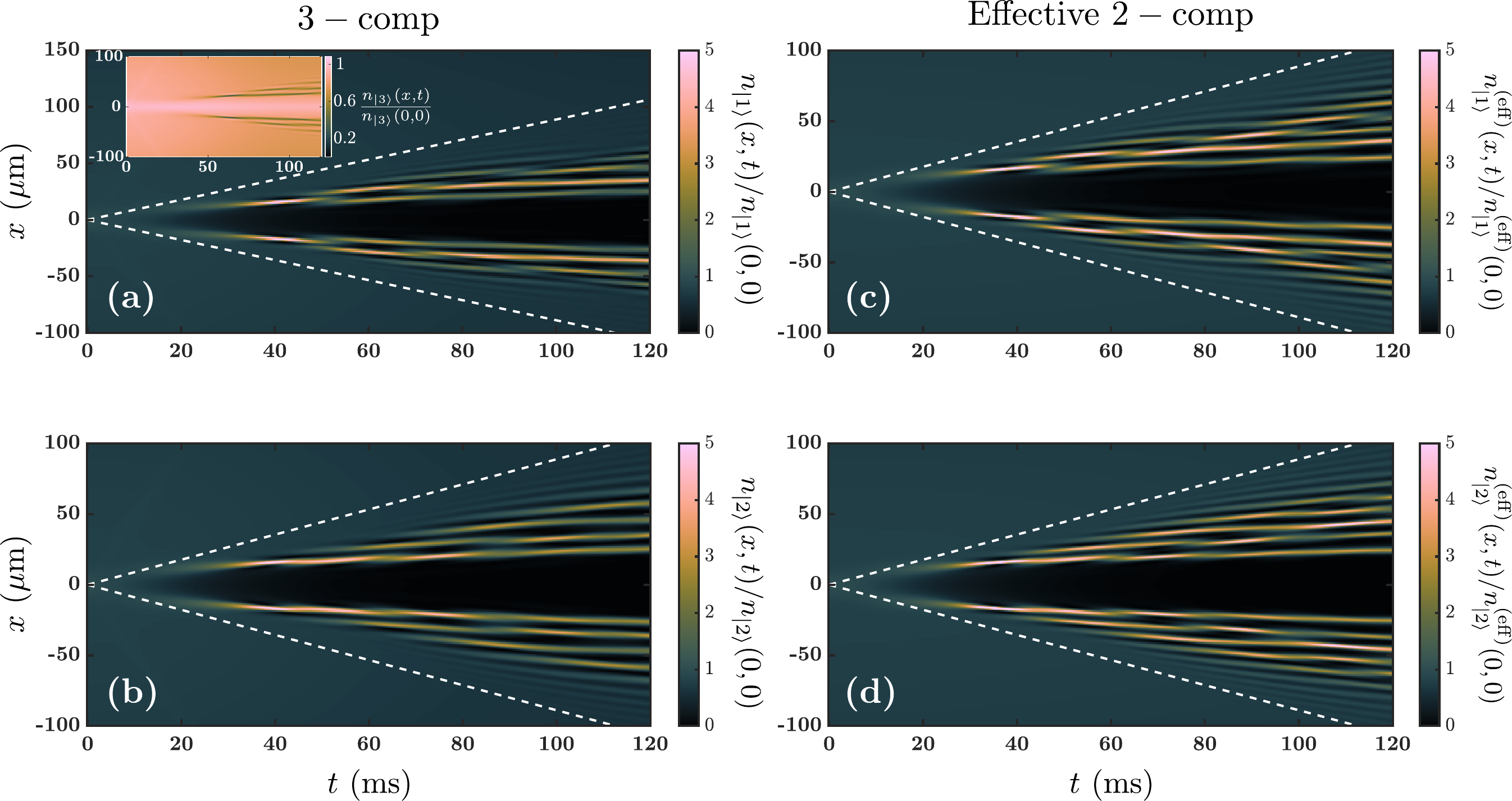}
    \caption{\textbf{Comparing the full three-component and reduced two-component models}. Spatiotemporal density evolution of the two minority components [$\ket{1} \equiv  \ket{1,-1}$ and $\ket{2} \equiv  \ket{1,0}$] pertaining to Setup S2, stemming from the (a), (b) full three-component system and (c), (d) the effective two-component reduction. The population imbalance is set to $f_{\rm{m}} = 5~\%$. Excellent agreement at short time scales is observed, where PSs occur, with the MI being more pronounced in the effective model. White dashed lines mark the MI envelope during the dynamics~\cite{mossman2024nonlinear}. Inset in (a) depicts the evolution of the $\ket{3} \equiv \ket{2,0}$ majority component, demonstrating the absence of  a PS in the case of $K^{(3)}_3=0$.}
    \label{Fig:Comparison_effective_model}
\end{figure}

\section{Supplementary Note 2: Validity of the effective model}  \label{Sec:Validity}

To assess the validity of the effective model for the three-component setting ($\mathcal{N}=3, k=1$), setup S2 is employed as a prototype since it is subject to strong inter- and intra-species MI.
Relevant investigations have been performed for the remaining setups discussed in the main text, where the agreement between the full three-component and the reduced two-component predictions is similar (and often even better) to the one described below.
For convenience, the pertinent predictions are provided within the one-dimensional (1D) GPEs.
Recall that in three-dimensions (3D) wave collapse occurs in the attractive interaction regime~\cite{Roberts_controlled_2001}.
The 1D GPEs are a reduction of Eq.~(1) in the main text, assuming the product wavefunction ansatz~\cite{kevrekidis_defocusing_2009} $\Psi_i(\boldsymbol{r},t) = \phi_i(x,t) \frac{e^{-\boldsymbol{r}^2_{\perp}/(2\ell^2_{\perp})}}{\ell_{\perp} \sqrt{\pi}}$, $i=1,2,3$, where $\boldsymbol{r}_{\perp}=(y,z)$ refer to the transverse coordinates, and $\ell_{\perp} = \sqrt{\hbar / (m \omega_{\perp})}$ is the oscillator length along the transverse directions.
Accordingly, the 1D GPEs for the three-component setting read
\begin{equation}
    i \hbar \partial_t \phi_i = \left[ -\frac{\hbar^2}{2m} \partial_x^2 + \frac{m}{2} \omega_x^2 x^2 + \sum_{j=1}^3 2\hbar \omega_{\perp} a_{ij} \abs{\phi_j}^2  \right] \phi_i, \quad i=1,2,3.
    \label{Eq:1D_reduction}
\end{equation}
This can be readily reduced to the two-component 1D GPEs (that will be used for the validation of the reduced model) in the case of $\phi_1=0$.
In fact, the latter represents the majority component, which is integrated out in the reduced two-component model.

\begin{figure}[t!]
    \centering
    \includegraphics[width=0.7\textwidth]{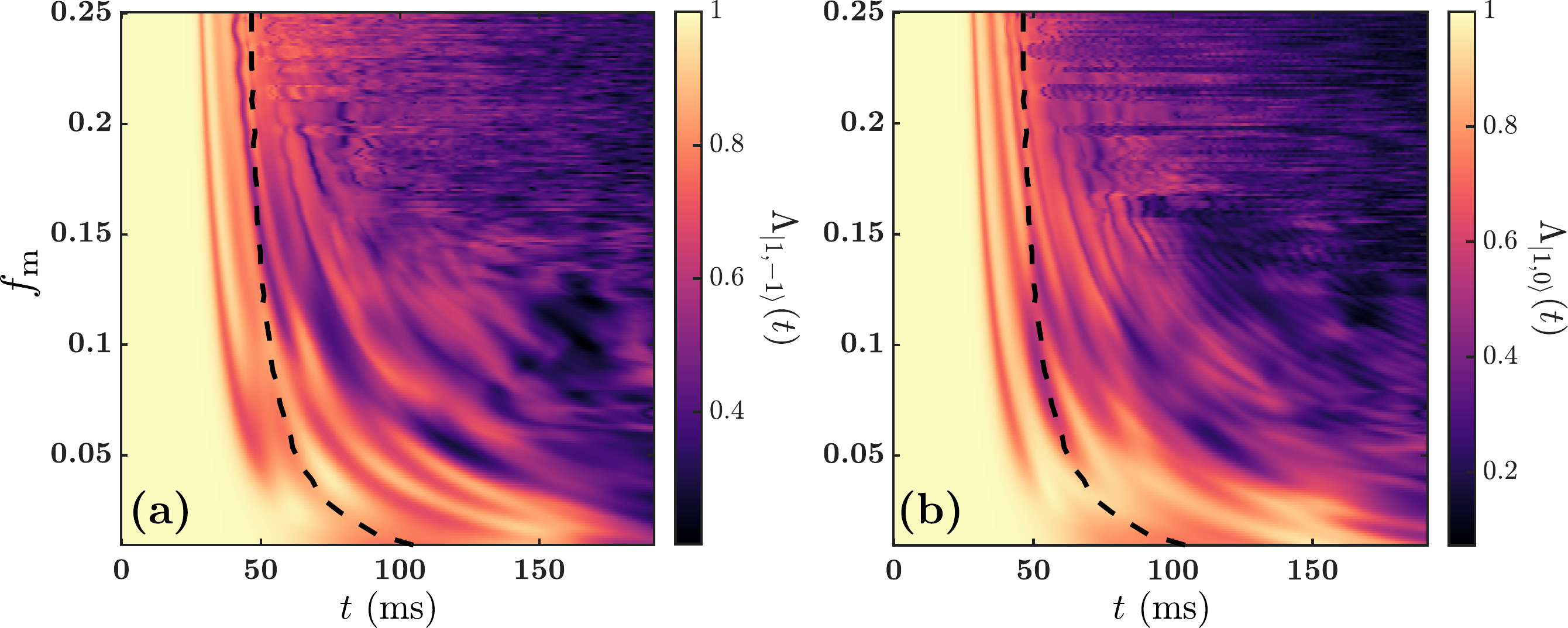}
    \caption{\textbf{Regions of validity of the effective reduction.} Comparison of the effective two-component system for the setup S2 with respect to the population imbalance, $f_{\rm{m}}$, in the course of the evolution. Excellent agreement above $90\%$ is observed until the  Peregrine formation, while later on more prominent deviations occur traced back to the highly oscillatory density structures caused by both inter- and intra-species MI [see also Supplementary Table~\ref{Tab:MI_characteristics}]. Panel (a) [(b)] presents the overlap measure (see main text) pertaining to the $\ket{1} \equiv | 1, -1 \rangle$ [$\ket{2} \equiv |1,0 \rangle$] minority component, as predicted by the three- and reduced two-component systems. Dashed lines denote the approximate time instants at which PSs are formed. 
    }
    \label{Fig:Overlaps}
\end{figure}

The density evolution of the minority components of Setup S2 following the experimental process outlined in the main text is presented in Supplementary Fig.~\ref{Fig:Comparison_effective_model}
for a small population imbalance ($f_{\rm{m}}= 5~\%$).
Overall qualitative agreement is observed between the same minority components as obtained from the full three-component and the effective model.
Specifically, in both cases twin PS nucleation takes place around $t \approx 40 ~ \rm{ms}$.
Small discrepancies arise after PS nucleation, where inter- and intra-species MI set in. 
The latter seed highly oscillatory density patterns propagating outwards. Notice here the sharp difference of the MI development, which does not encompass a central density portion as the one shown in Fig.~2(b3), (b5) of the main text.
This highlights the importance of three-body losses for capturing the experimental observations. 
The MI envelope is captured to an excellent degree by the prediction of Ref~\cite{mossman2024nonlinear}, referring to two-component settings, see white dashed lines in Supplementary Fig.~\ref{Fig:Comparison_effective_model}. 
Naturally, the twin PSs are characterized by a large amplitude as compared to their background. 
This way, they locally become comparable to the majority component [see inset in Supplementary Fig.~\ref{Fig:Comparison_effective_model}(a)], thus
signaling the breakdown of the effective description, see also Supplementary Fig.~\ref{Fig:Overlaps}.
Notice here the lack of PS generation in the majority component, when excluding three-body losses.

The aforementioned agreement/breakdown of the effective picture is cleanly visualized by inspecting the overlap of the effective, $\phi^{(\rm{eff})}_{\rm{m}}$, (emanating from the reduced model) with the actual, $\phi_{\rm{m}}$, (stemming from the full three-species system) minority components~\cite{Erdmann_phase_2019,Mistakidis_correlation_2018,Bandyopadhyay_dynamics_2017,Jain_quantum_2011},

\begin{equation}
    \Lambda_{\rm{m}}(t) = \frac{  \left[  \int dx ~ \abs{\phi_{\rm{m}}(x,t)}^2 \abs{\phi^{(\rm{eff})}_{\rm{m}}(x,t)}^2  \right]^2  }{ \left[  \int dx ~ \abs{\phi_{\rm{m}}(x,t)}^4 \right]  \left[  \int dx ~ \abs{\phi^{(\rm{eff})}_{\rm{m}}(x,t)}^4 \right]   },
    \label{Eq:Overlaps}
\end{equation}
presented in Supplementary Fig.~\ref{Fig:Overlaps} for the $\ket{1,-1}$ [panel (a)] and $\ket{1,0}$ [panel (b)] hyperfine level of Setup S2.
Perfect overlap ($\Lambda_{\rm{m}}(t) \simeq 1$) is observed only during the initial evolution times before the twin PSs nucleation (marked by the dashed lines in Supplementary Fig.~\ref{Fig:Overlaps}).
Around PSs formation $\Lambda_{\rm{m}}(t) \in [0.8,1]$, signifying that deviations from the full system dynamics are becoming more prominent for longer evolution times.
Let us note, however, that $\Lambda_{\rm{m}}(t)$ is a strict quantitative measure, and the overall dynamical response is adequately captured qualitatively by the reduced model deep in the evolution [see also Supplementary Fig.~\ref{Fig:Comparison_effective_model}].

\section{Supplementary Note 3: Dynamical emergence of vector and Twin Peregrines }  \label{Sec:Spacetimes}

We demonstrate the dynamical PS nucleation for Setups S4-S6 discussed in the main text, following the experimental radio frequency process, see also Fig.~3 in the main text.
For simplicity, our results pertain to the NPGPE in the absence of three-body losses ($K^{(j)}_3 = 0$). 
Recall that S4 involves the hyperfine states $\ket{1} \equiv  \ket{1,-1}$, $\ket{2} \equiv \ket{1,0}$, and $\ket{4} \equiv  \ket{2,-2}$.
Moreover, S5 pertains to the hyperfine levels $\ket{5} \equiv  \ket{1,1}$, $\ket{2} \equiv \ket{1,0}$ and $\ket{1} \equiv \ket{1,-1}$, while S6 contains the states $\ket{4} \equiv \ket{2,-2}$, $\ket{6} \equiv \ket{2,-1}$ and $\ket{7} \equiv  \ket{2,1}$.
For our simulations here we consider majority-minority imbalance $f_{\rm{m}} = 10~\%$ with the majority and minority states for each setup summarized in Table~I in the main text along with the associated 3D scattering lengths of the full three-component mixture.

\begin{figure}
    \centering
    \includegraphics[width=0.7\textwidth]{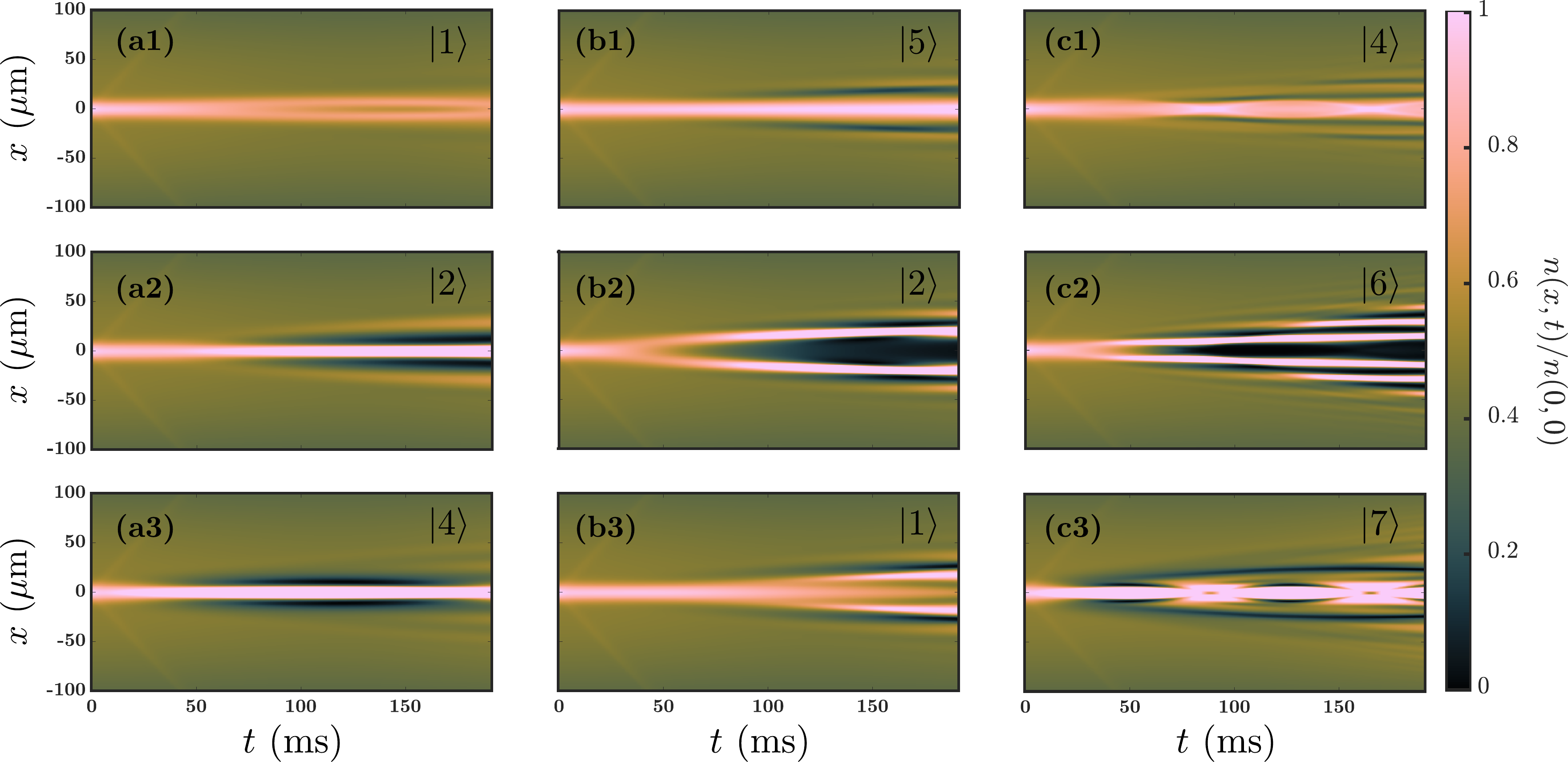}
    \caption{
    \textbf{Vector PS formation in three-component BECs}. Spatiotemporal density evolution of the three components for Setups S4 [(a1)-(a3)], S5 [(b1)-(b3)] and S6 [(c1)-(c3)], obtained for the radio frequency process emulated within the NPGPE. Collocated PS structures emerge in S4 atop $\ket{2}$ and $\ket{4}$ minority states, while in S5 and S6 a twin PS is observed in $\ket{2}$ and $\ket{6}$ minorities  respectively (see also text). 
    The majority components are depicted in (a1), (b1) and (c1) for the three setups. In all cases the majority-minority population imbalance corresponds to $f_{\rm{m}} = 10~\%$.}
    \label{Fig:Spatiotemporal_Supp}
\end{figure}

Supplementary Figure~\ref{Fig:Spatiotemporal_Supp} presents the density evolution of the individual hyperfine states for all three setups.
Focusing on S4, one PS appears per component due to the inter MI [Supplementary Fig.~\ref{Fig:Spatiotemporal_Supp}(a2), (a3)], see Supplementary Table~\ref{Tab:MI_characteristics}. The vector PSs emerge at slightly different times.
This behavior becomes even more prominent in the extreme majority-minority imbalance, corresponding to $f_{\rm{m}} \approx 1 ~\%$ (not shown), and it is attributed to the immiscibility of the minority components predicted by the effective two-component reduction. 
On the other hand, the majority component residing at the trap center gradually develops a density dip at the location of the vector PS, see Supplementary Fig.~\ref{Fig:Spatiotemporal_Supp}(a1).
Turning to S5 [S6], a twin PS appears in the $\ket{2}$, Suplementary Fig.~\ref{Fig:Spatiotemporal_Supp}(b2) [$\ket{6}$, Supplementary Fig.~\ref{Fig:Spatiotemporal_Supp}(c2)] state due to the intra-species MI character in this minority component, while remaining predominantly immiscible with the $\ket{5}$, Supplementary Fig.~\ref{Fig:Spatiotemporal_Supp}(b1) [$\ket{4}$, Supplementary Fig.~\ref{Fig:Spatiotemporal_Supp}(c1)] and $\ket{1}$, Supplementary Fig.~\ref{Fig:Spatiotemporal_Supp}(b3) [$\ket{7}$, Supplementary Fig.~\ref{Fig:Spatiotemporal_Supp}(c3)].
Notice that the twin PS structure in S6 emerges faster compared to S5 due to the larger intracomponent attraction driving the intra MI. 
Consequently, MI sets in faster in S6 in contrast to S5, manifested through the filamentation of the density of the $\ket{6}$ minority component.

\section{Supplementary Note 4: Non-polynomial GPE reduction}  \label{Sec:quasi-1D}

Due to the experimentally used trap aspect ratios, the three-component mixture is mostly kinematically constrained across the elongated direction, i.e. it is a quasi-1D system.
In this sense, a lower dimensional description can be employed~\cite{Salasnich_effective_2002}, serving as a useful guide for the rich phenomenology of vector PSs in three-component settings.

The starting point is the action pertaining to the three-component system in 3D,

\begin{equation}
    S[\Psi_j] = \int dt\,d^3\boldsymbol{r} ~ \Bigg\{  \sum_{i=1}^3 \left[  -i\frac{\hbar}{2} \left( \Psi_i^*  \partial_t \Psi_i  - \Psi_i  \partial_t \Psi_i^* \right) + \frac{\hbar^2}{2m} \abs{\nabla \Psi_i}^2 + V(\boldsymbol{r}) \abs{\Psi_i}^2 +\frac{g_{ii}}{2} \abs{\Psi_i}^4 \right] + \sum_{i>j} g_{ij} \abs{\Psi_i}^2 \abs{\Psi_j}^2  \Bigg\}.
    \label{Eq:Action}
\end{equation}
Here, $\Psi_i(\boldsymbol{r},t)$, $i=1,2,3$, denote the 3D wavefunctions pertaining to the three different hyperfine states, and $V(\boldsymbol{r})$ is the trapping potential, decomposed as $V(\boldsymbol{r}) = U(x) + \frac{1}{2} m \omega_{\perp}^2 \boldsymbol{r}^2_{\perp}$. The potential along the axial direction, $U(x)$, can be in principle arbitrary.
The quasi-1D description relies on the following product ansatz for the axial and transverse profiles,

\begin{equation}
    \Psi_i(\boldsymbol{r},t) = \varphi_i(x,t) \frac{e^{-\boldsymbol{r}^2_{\perp}/(2\sigma^2)}}{\sigma \sqrt{\pi}},
    \label{Eq:Quasi_1D_ansatz}
\end{equation}
where we have assumed that the transverse profile is the same for all components. 
The width $\sigma$ is a variational parameter, determined from the Euler-Lagrange equations,
\begin{equation}
    \sigma^4 = \ell_{\perp}^4 \Bigg\{  1 + \frac{\sum_{i \geq j} 2 a_{ij} \abs{\varphi_i}^2 \abs{\varphi_j}^2}{\sum_{i=1}^3 \abs{\varphi_i}^2}  \Bigg\}.
    \label{Eq:Width}
\end{equation}
Note that the width is coupled to the axial profiles, $\varphi_i$. 
The corresponding quasi-1D equations of motion are extracted from the Euler-Lagrange equations of the action described by Eq.~\eqref{Eq:Action},

\begin{equation}
    i\hbar \frac{\partial \varphi_i}{\partial t} = \left[  -\frac{\hbar^2}{2m} \partial_x^2 + U(x) + \sum_{j=1}^3 \frac{g_{ij}}{2\pi \sigma^2} \abs{\varphi_j}^2 +\frac{\hbar^2}{2m \sigma^2} + \frac{m \omega_{\perp}^2}{2} \sigma^2  \right] \varphi_i, \quad i=1,2,3.
\end{equation}
The latter are the non-polynomial GPEs (NPGPEs) for the three-component system, generalizing this way earlier findings concerning single~\cite{Salasnich_effective_2002} and two-component~\cite{salasnich2009generalized} mixtures.

\putbib[Three_comps]

\end{bibunit}

\end{document}